\documentclass[useAMS,usenatbib]{mn2e}
\usepackage{latexsym}
\usepackage{amsmath}
\usepackage{amssymb}
\usepackage{float}
\usepackage{subcaption}
\usepackage{xcolor}
\usepackage{caption}
\usepackage{graphicx}
\title[On filament fragmentation and the impact of ambient environment on it]{On filament fragmentation and the impact of ambient environment on it}
\author[S. V. Anathpindika and Di Francesco, James]{S. V. Anathpindika$^{1,}$ $^{2}$\thanks{E-mail: sumed$\_$k@yahoo.co.in (SVA)} and Di Francesco, J$^{3}$\\
  $^{1}$Indian Institute of Technology, Kharagpur, West Bengal, India\\
  $^{2}$University Observatory M$\ddot{\mathrm{u}}$nchen, Schneirstrasse 1, 81679, M$\ddot{\mathrm{u}}$nchen, Germany \\
  $^{3}$National Research Council of Canada, Herzberg, Astronomy \& Astrophysics Research Centre, \\
  5071 West Saanich Road, Victoria (BC), Canada V9E 2E7 \\}
\begin{document}

\date{Accepted 0000 December 00. Received 00 December 00; in original form 0000 October 00}

\pagerange{\pageref{firstpage}--\pageref{lastpage}} \pubyear{2002}

\maketitle

\label{firstpage}

\begin{abstract}
Filaments are crucial intermediaries in the star formation process. Recent observations of filaments show that - \textbf{(i)} a number of them are non-singular entities, and rather a bundle of velocity coherent fibres, and \textbf{(ii)} while a majority of filaments spawn cores narrower than their natal filaments, some cores are broader. We explore these issues by developing hydrodynamic simulations of an initially sub-critical individual filament that is allowed to accrete gas from its neighbourhood and evolves under self-gravity. Results obtained here support the idea that fibres form naturally during the filament formation process. We further argue that the ambient environment, i.e., the magnitude of external pressure, and not the filament linemass alone, has bearing upon the morphology of its evolution. We observe that a filament is susceptible to the \emph{sausage}-type instability irrespective of the external pressure. The fragments, however, are pinched in a filament experiencing pressure comparable to that in the Solar neighbourhood ($\sim 10^{4}$ K cm$^{-3}$). By contrast, fragments are broad and spherical - having density profiles similar to that of a stable Bonnor - Ebert sphere - when the filament experiences a higher pressure, typically $\ge 10^{5}$ K cm$^{-3}$, but $\le 10^{6}$ K cm$^{-3}$). The filament tends to rupture at even higher external pressure ($\ge 10^{7}$ K cm$^{-3}$). These observations collectively mean that star formation is less efficient with increasing external pressure.
\end{abstract}


\section{Introduction}
Filaments play a pivotal role in the star-formation process. For instance, dense gas within molecular clouds ({\small MC}s) often appears filamentary on the plane of the sky (e.g., Schneider \& Elmegreen 1979, Myers 2009, Schisano \emph{et al.} 2014, Marsh \emph{et al.} 2016). 
Indeed, \emph{Herschel} continuum surveys of {\small MCs} established the ubiquity of dense filaments (e.g., Andr{\'e} \emph{et al.} 2010, Jackson \emph{et al.} 2010, Men'shchikov \emph{et al} 2010, Molinari \emph{et al.} 2010, see also the review by Andr{\'e} \emph{et al.} 2014 - to list a few), and prestellar cores appear aligned with them much like pearls in a necklace. In the Aquila Rift cloud, roughly three quarters of prestellar cores are found in dense, super critical filaments, i.e., those with linemasses (masses per unit length) in excess of the critical value, $M_{crit}\equiv\frac{2a_{0}^{2}}{G}\sim$ 16.4 M$_{\odot}$ pc$^{-1}$; $a_{0}\sim$ 0.2 km s$^{-1}$, being the sound speed for molecular gas at 10 K (e.g., K$\ddot{\mathrm{o}}$nyves \emph{et al.} 2015). Prestellar cores inherit their physical characteristics such as the density and temperature profile from the natal filament and so, the further evolution of cores must fundamentally depend on the properties of the filament itself (e.g., Keto \& Burkert 2014). \\ \\
Observations of filaments in the local neighbourhood support an evolutionary sequence in which filaments gradually evolve from a thermally sub-critical state (i.e., with linemass less than $M_{crit}$) to a super-critical state (i.e., with linemass greater than $M_{crit}$) via gravitational contraction, see e.g., Arzoumanian \emph{et al.} (2013). These latter authors also report that the velocity dispersion, $\sigma_{gas}$, within filaments correlates with the linemass, $M_{l}$, and the column density, $N_{H_{2}}$ as $\sigma_{gas}\propto M_{l}^{0.36}$ and $\sigma_{gas}\propto N_{H_{2}}^{0.5}$, respectively.  Filaments also appear to be universally characterised by an inner width on the order of $\sim$0.1 pc (e.g., Arzoumanian \emph{et al.} 2011, 2019; Koch \& Rosolowsky 2015). In the paradigm of turbulence-driven filament formation ({\small TDFF}), this length signifies the scale below which turbulent gas must become subsonic. \\ \\
In the {\small TDFF} paradigm, it is possible for a filament to split into mutually intertwined fibres while accreting gas from the natal cloud, and for some of these fibres to fragment and form cores - the so-called \emph{fray and fragment} scenario. Indeed, the filament {\small B211/213} in the Taurus {\small MC} appears to be consistent with this scenario, as reflected by the velocity coherent substructure seen along its length (Hacar \emph{et al.} 2013). Other filaments such as the {\small L1517 MC} exhibit oscillatory line-of-sight velocity patterns, implying filament fragmentation leading to core formation (Hacar \& Tafalla 2011). This behaviour, however, is not universal as other filaments exhibit more complex velocity patterns (Tafalla \& Hacar 2015). {\small TDFF} has been routinely observed in numerical simulations of turbulent gas (e.g., Padoan \emph{et al.} 2001, MacLow \&  Klessen 2004, Federrath \emph{et al.} 2010, Smith \emph{et al.} 2014). \\ \\
In the {\small TDFF} picture, filaments are associated with stagnation regions where turbulent elements of gas converge, causing energy to dissipate. Furthermore, filaments and cores often appear to form nearly simultaneously (e.g., Gong \& Ostriker 2011, G{\'o}mez \& V{\'a}zquez-Semadeni 2014). The appearance of velocity coherent fibres along an individual filament that was allowed to accrete gas from its surroundings was demonstrated numerically by Clarke \emph{et al.} (2017) by modelling a typical filament as an isothermal cylinder, superposed with a turbulent velocity field. Formation of filaments and their associated fibres were also seen in simulations of a turbulent {\small MC}, e.g. by, Smith \emph{et al.} (2014). While the Clarke \emph{et al.} simulations lend support to the \emph{fray and fragment} scenario, other simulations of turbulent clouds such as those by Smith \emph{et al.} do not disambiguate between whether the filaments themselves split into fibres or a bundle of fibres gather to form a filament. \\\\
Hydrodynamic simulations have also been used to
look into modes of filament fragmentation. Gritschneder \emph{et al.} (2017) showed that a  small initial sinusoidal perturbation on a filament with $f_{cyl}=0.5$ \footnote{$f_{cyl}=\frac{M_{l}}{M_{l_{crit}}}$}, could amplify with time and fragment the filament, the so-called \emph{geometrical fragmentation} of a filament. Similarly, Heigl \emph{et al.} (2019) argued that the initial linemass possibly determines the morphology of cores that eventually form in a filament. They showed that a filament with initially $f_{cyl}=0.8$ (transcritical)  contracted and became susceptible to a Jeans-like \emph{compressional} instability before fragmenting to produce cores that were \emph{pinched}, i.e., the core size was smaller than the parent filament width. By contrast, an initially sub-critical filament ($f_{cyl}=0.2$) contracted and became susceptible to a \emph{sausage-type} instability before fragmenting to produce broad cores, i.e., the core-size was greater than the parent filament width. Indeed, in his analytic work on the stability of a filament, Nagasawa (1987) found that a filament of radius smaller than its scale height became unstable to the \emph{sausage-like} instability. Conversely, when the radius is larger than its scale height the filament became unstable to a Jeans-like \emph{compressional} instability.\\\\
The deformation or the \emph{sausage-type} instability occurs when the volume energy due to the external pressure dominates the energy budget of a filament. In this case, even a slight decrease in the volume energy due to a deformation along the length of a filament is sufficient to render it unstable and a \emph{sausage-type} instability ensues. On the other hand, when the volume energy due to external pressure is relatively small such that the equilibrium configuration of the filament is governed mainly by a balance between the internal energy (purely thermal in Nagasawa's work) and the gravitational energy, the filament becomes unstable to local density enhancements, the so-called compressional instability. In their analytic / numerical work, Inutsuka \& Miyama (1992) reported the formation of spherical (broad) cores in an initially transcritical filament ($f_{cyl}$ = 1) experiencing a relatively large external pressure. In their work initially super-critical filaments  confined by a relatively small external pressure collapsed radially leading to pinched cores. In light of the recent findings reported by Heigl \emph{et al.}, however, it appears that the \emph{broad cores} in their work are a manifestation of the deformation instability while the \emph{pinched} cores result from the compressional instability.\\\\
Although these works clearly show the possible variety in the modes of filament evolution, understanding the physical conditions under which a filament might exhibit any particular mode continues to evade us. The work by Heigl \emph{et al.}, Gritschneder \emph{et al.}, or indeed the previous work by Inutsuka \& Miyama (1992, 1997) motivates a picture in which the filament evolution depends exclusively on its linemass quantified by the parameter, $f_{cyl}$. These works, however, do not discuss the impact of physical conditions that could be crucial in determining the difference in the mode of filament evolution. We therefore do not fully appreciate the physical conditions under which a filament is likely to yield pinched or broad cores. To reconcile better the initial conditions for star formation, in this paper we  probe further the evolution of filaments and the origin of important physical properties such as the characteristic width, the temperature profile and the propensity to fragment, as well as the impact of the ambient environment on these properties. Much of the recent numerical work that focused on fragmentation of filaments and the formation of cores in them, has generally assumed isothermal conditions. \\\\
In this paper, we discuss results from realisations in which gas is allowed to cool dynamically so that the evolution of the temperature profile of a filament initially supported by turbulence can be probed. Besides exploring the modes of filament fragmentation, we also derive density probability distribution functions ({\small PDFs}) for filaments just as they begin to fragment. We also explore other important correlations such as those between the gas velocity dispersion and the column density and the filament linemass. The plan of this paper is as follows. We describe the numerical code and the initial conditions used here in \S 2. Results, the morphology of filament fragmentation, and typical physical diagnostics deduced here are described in \S 3. The results are discussed in \S 4 before conclusions are summarised in \S 5.
\section[]{Numerical Method and Initial conditions}
\subsection{Some basic aspects of filament evolution}
A filament is often idealised as an isothermal cylinder of uniform density and is characterised by its linemass, $M_{l}$. The critical value of linemass is $M_{l_{crit}} = \frac{2a_{0}^{2}}{G}$ = 16.4 M$_{\odot}$ pc$^{-1}$; $a_{0}$ = 0.2 km/s at 10 K. A stability analysis of a self-gravitating cylinder by Nagasawa (1987) showed that the fastest growing wavenumber in it was, $k_{fast}$ = 0.284$(4\pi \mathrm{G}\rho_{c})^{0.5}/a_{0}$, corresponding to a fragmentation length-scale, 
\begin{equation}
\lambda_{frag} = \frac{2\pi}{k_{fast}} = 22.1\frac{a_{0}}{(4\pi \mathrm{G}\rho_{c})^{0.5}}.\end{equation}
The growth rate of this mode was, $\omega_{max}$ = 0.339$(4\pi \mathrm{G}\rho_{c})^{0.5}$.  
\\ \\
Inutsuka \& Miyama (1992) further showed that a quasi-stable or a transcritical cylinder ($M_{l}\sim M_{l_{crit}}$) is prone to fragmentation and can break-up into smaller fragments. The length of the fastest growing unstable mode is $\sim 8H$, where
\begin{equation}
H = \frac{a_{0}}{(4\pi G\rho_{c})^{0.5}},
\end{equation}
 is the scale-height for a filament. A super critical cylinder, i.e., $M_{l}\gg M_{l_{crit}}$, on the other hand should collapse radially without much fragmentation. Besides the fragmentation length-scale,  Fischera \& Martin (2012) showed analytically the dependence of filament properties such as its central density, radius, and column density on the magnitude of external pressure, $P_{ext}$. They further showed that the central density of a cylinder varied as $P_{ext}^{0.5}$ while its FWHM varied as $P_{ext}^{-0.5}$. It is in light of these analytic predictions that we further study the evolution of filaments in this paper.
\begin{table*}
 \centering
 \begin{minipage}{160mm}
  \caption{Physical parameters for realisations.}
  \begin{tabular}{@{}lllll@{}}
  \hline
   Serial & Magnitude of & Gas Temperature & Velocity & Mach \\
   Number & Ext. press. & (T$_{gas}$/[K])& [Gas (V$_{fil}$), & Nos. \\
          & &Temperature of externally  & External medium (V$_{acc}$),&\\
          & & confining {\small ICM}    &                             & \\
          & $\frac{(P_{ext}/k_{B})}{[\mathrm{K} \ \mathrm{cm}^{-3}]}$ &(T$_{ext}$/[K]) & Intercloud medium (V$_{ext}$)] &  [km/s]\\
\hline
   \multicolumn{5}{c}{M$_{fil}$ = 41.6 M$_{\odot}$, L$_{fil}$ = 5 pc, r$_{fil}$ = 0.2 pc, n$_{avg}$= 619 cm$^{-3}$, f$_{cyl}$= 0.5 \footnote{f$_{cyl}$ = $\frac{M_{l}}{M_{l_{crit}}}$}} \\
 \hline
 1 &$2\times10^{4}$ & T$_{gas}$=11 & V$_{fil}$=0.14, V$_{acc}$=0.72  & $\mathcal{M}_{gas}$ = 0.7\\
   & &  T$_{ext}$=197.1 & V$_{ext}$=0.66 & $\mathcal{M}_{ext}$=0.8 \\
 \hline
  2 &$2\times10^{5}$ & T$_{gas}$=14.4 & V$_{fil}$=0.71, V$_{ext}$=2.35  & $\mathcal{M}_{gas}$ = 3.2\\
   & &  T$_{ext}$=134.5 & V$_{ext}$=3.27 & $\mathcal{M}_{ext}$=4.8 \\
 \hline
 3 &$2\times10^{6}$ & T$_{gas}$=14.6 & V$_{fil}$=2.35, V$_{acc}$=7.46  & $\mathcal{M}_{gas}$ = 10.5\\
   & &  T$_{ext}$=223 & V$_{ext}$=10.52 & $\mathcal{M}_{ext}$=12.0 \\
 \hline
 4 &$2\times10^{7}$ & T$_{gas}$=36.8 & V$_{fil}$=7.46, V$_{acc}$=23.6  & $\mathcal{M}_{gas}$ = 21.0\\
   & &  T$_{ext}$=560.3 & V$_{ext}$=33.4 & $\mathcal{M}_{ext}$=24.0 \\
 \hline
 5 & \multicolumn{3}{c}{Same as Case 2, but isothermal equation of state (T$_{gas}$=14.4 K, T$_{\small ICM}$=134.5 K)}\\
 \hline
 \multicolumn{5}{c}{M$_{fil}$ = 58.7 M$_{\odot}$, L$_{fil}$ = 5 pc, r$_{fil}$ = 0.2 pc, n$_{avg}$= 866 cm$^{-3}$, f$_{cyl}$= 0.7} \\
 \hline
 6 &$2\times10^{4}$ & T$_{gas}$=11 & V$_{fil}$=0.04, V$_{acc}$=0.6  & $\mathcal{M}_{gas}$ = 0.2\\
   & &  T$_{ext}$=141.0 & V$_{ext}$=0.55 & $\mathcal{M}_{ext}$=0.8 \\
\hline
\multicolumn{5}{c}{M$_{fil}$ = 25.2 M$_{\odot}$, L$_{fil}$ = 5 pc, r$_{fil}$ = 0.2 pc, n$_{avg}$= 371 cm$^{-3}$, f$_{cyl}$= 0.3} \\
\hline
 7 &$2\times10^{4}$ & T$_{gas}$=11 & V$_{fil}$=0.24, V$_{acc}$=0.94  & $\mathcal{M}_{gas}$ = 1.2\\
   & &  T$_{ext}$=269.4 & V$_{ext}$=0.96 & $\mathcal{M}_{ext}$=1.0 \\
\hline
\end{tabular}
\end{minipage}
\end{table*}
\subsection{Numerical Method} The realisations we describe in this paper were developed using the Smoothed Particle Hydrodynamics ({\small SPH}) code {\small SEREN}. This code is well tested (Hubber \emph{et al.} 2011). It includes the artificial thermal conductivity prescribed by Price (2008) to promote mixing between {\small SPH} particles in regions that exhibit a density gradient and thus avoids problems related to resolution of discontinuities and shocks. We include the carbon and hydrogen chemistry described by Bate \& Keto (2015). Besides heating due to the release of gravitational potential, other contributors responsible for heating a gas particle are cosmic rays, photoelectric heating, and heating due to the formation of molecular hydrogen, are all included in these calculations and quantified by the respective heating functions $\Gamma_{cr}$, $\Gamma_{pe}$ and $\Gamma_{H_{2},g}$. Of these, $\Gamma_{cr}$ and $\Gamma_{pe}$ are held constant while $\Gamma_{H_{2},g}$ is calculated by explicitly solving for the change in hydrogen fraction at each time step. \\\\
Similarly, factors contributing to gas cooling are emissions due to electron recombination, singly ionised oxygen and the carbon fine-structure cooling characterised by the respective cooling functions $\Lambda_{rec}$, $\Lambda_{OI}$ and $\Lambda_{C^{+}}$. We also include cooling due to molecular line emission characterised by the cooling function, $\Lambda_{line}$. The thermal interaction between gas and dust is accounted for by including the rate for gas-dust interaction, $\Lambda_{gd}$, which may contribute to net heating or cooling depending on whether the dust is warmer or cooler than the gas, respectively. Respective equations of thermal balance for gas and dust are then iteratively solved to calculate their equilibrium temperature. {\small SPH} particles denser than a certain threshold ($10^{7}$ cm$^{-3}$) were replaced with dead particles i.e., \emph{sink} particles, but we are not interested here in collating statistics of putative star particles and therefore do not follow their evolution in these realisations. Instead, we terminated the calculations after 2 - 3 sink particles formed in the filament in each case. We employed an adaptive smoothing length so that each {\small SPH} particle had exactly 50 neighbours to minimise inaccuracies in the estimation of the density field (e.g. Attwood \emph{et al.} 2007).
\begin{figure*}
  \vspace{10pt}
  \includegraphics[angle=0,width=\textwidth]{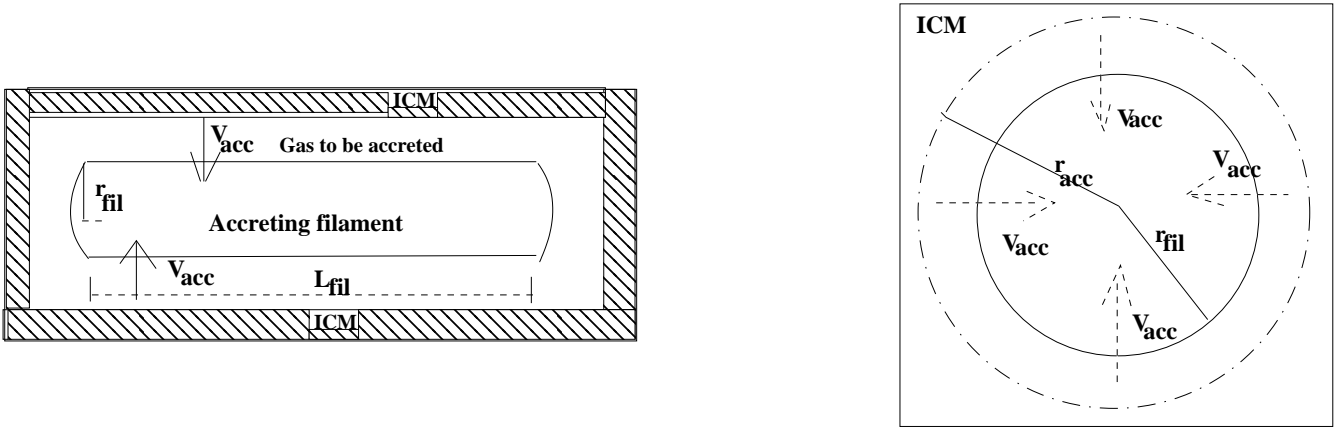}
  \caption{On the left is a schematic \small{2-D} illustration of the initial set-up for the model filament represented as a cylinder of length, $L_{fil}$, and radius, $r_{fil}$, which is enveloped by a jacket of gas to be accreted by the filament. The filament and its jacket are confined by the \small {ICM} and the entire assembly is placed in a periodic box. A schematic of the cross-section of this assembly is shown on the right.}
\end{figure*}
\subsection{Initial conditions}
The model filament probed in this paper is a finite turbulence-supported cylinder - unlike the infinitely long cylinder assumed by Clarke \emph{et al.} (2017) - that has initially uniform density and which is allowed to accrete gas from its surroundings. The filament initially has an aspect ratio of $\sim$25 with its axis aligned along the $x-$direction and gas was allowed to accrete along the directions orthogonal to its axis, i.e., along the $y$- and $z$- axes. The filament is characterised by initial mass, $M_{fil}$, length, $L_{fil}$, radius $r_{fil}$, average initial density, $\rho_{fil}$ and internal pressure, $P_{int}$. Gas to be accreted by the filament was initially placed in an envelope that had radial and axial thickness $dr$ and $dl$ respectively, so that the radius and length of this envelope were $r_{acc} = r_{fil} + dr$, and $l_{acc} = l_{fil} + dl$. Both $dl$ and $dr$ were chosen to be on the order of twice the average initial {\small SPH} smoothing length, $h_{avg}$. The filament and the envelope of gas to be accreted by it were assembled such that their interface was initially in approximate pressure balance. Other physical parameters associated with the gas to be accreted are its mass, $M_{acc}$, the magnitude of velocity associated with it, $V_{acc}$, the temperature, $T_{acc}$, and the average initial density, $\rho_{acc}$. Finally, the filament and its envelope were jacketed with another layer of particles representing the intercloud medium ({\small ICM}) to simulate the external pressure. The entire set-up, i.e., the filament, the envelope of gas to be accreted and the {\small ICM} was enclosed in a periodic box. The cartoons in Fig. 1 show a schematic representation of the computational domain. Realisations were developed by employing the full periodic boundary conditions.\\\\
Gas to be accreted by the filament has density an order of magnitude lower than that in the filament; $X_{acc} \equiv \frac{\rho_{acc}}{\rho_{fil}}$ = 0.1, in all test realisations. The cylinder is initially supported by a combination of turbulence having magnitude of velocity, $V_{fil}$, and thermal pressure. Thus we have,
\begin{equation}
P_{int}\equiv \rho_{fil}(a_{gas}^{2} + V_{fil}^{2}) = P_{acc}\equiv \rho_{acc}(a_{acc}^{2} + V_{acc}^{2}).
\end{equation}
Here $a_{acc}$ is the sound speed corresponding to the temperature, $T_{acc}$, and the corresponding pressure of the gas to be accreted by the filament, $P_{acc}$. Likewise, $T_{gas}$ and $a_{gas}$ are, respectively. the temperature of the gas and the corresponding sound speed.
The turbulent velocity field in the cylinder is characterised by the Mach number, $\mathcal{M}_{fil}$. Since we assume that gas to be accreted has the same initial temperature as the gas in the cylinder, Eqn. (3) above can be simply flipped to obtain the velocity, $V_{acc}$. Finally, the pressure due to the externally confining {\small ICM} is,
\begin{equation}
P_{ext} = P_{int} + P_{acc} \equiv \rho_{ext}(a_{ext}^{2} + V_{ext}^{2}).
\end{equation}
As usual, $a_{ext}$ is the sound speed of gas representing the {\small ICM} and $V_{ext}$ is the magnitude of its velocity; $X_{ext} \equiv \frac{\rho_{ext}}{\rho_{fil}}$ = 0.1. As with the gas in the filament, the externally confining {\small ICM} is also characterised by its Mach number, $\mathcal{M}_{ext}$. The respective turbulent velocity fields in the cylinder and the {\small ICM} were modelled with a fairly steep power spectrum ($\sim k^{-4}$) which is consistent with the suggestions of Burkhart \emph{et al.} (2010) for diffuse, optically thin gas at intermediate Mach numbers. \\\\
As can be seen from Table 1, we have restricted our choice of the external pressure, $P_{ext}$ to between $10^{4}$ K cm$^{-3}$ - 10$^{7}$ K cm$^{-3}$. This choice of pressure is representative of the range of pressure reported at different locations in the Galactic disk. Keeping track of the number of parameters listed in Table 1 could, on the face of it, appear cumbersome and so, we consider one particular case, say Case 3 here for illustrative purposes. For the filament in this case to be confined by an {\small ICM} pressure of magnitude, $\frac{P_{ext}}{k_{B}} = 2\times 10^{6}$ K cm$^{-3}$, we begin by assuming the initial pressure within the filament to be, $\frac{P_{int}}{k_{B}} = 10^{6}$ K cm$^{-3}$, because the interface between the filament and the external gas to be accreted by it is initially assumed to be in approximate equilibrium. Thus the combined pressure due to the filament and the envelope of gas to be accreted is $2\times 10^{6}$ K cm$^{-3}$, which must be equal to the external pressure by design (see Col. 2 of Table 1). The density contrast, $X_{acc}$, and $X_{ext}$, as noted above, are both fixed at 0.1. The average initial gas density within the filament is calculated simply using its dimensions and the gas is assumed to be initially cold with the choice of gas temperature, $T_{gas}$, listed in column 3 of Table 1. Consequently, we are left with only two other parameters to be chosen viz., $\mathcal{M}_{gas}$ and $\mathcal{M}_{ext}$ in respectively Eqns. (3) and (4). They are chosen such that the desired pressure is obtained. The sound speed is converted into the turbulent velocity dispersion in the usual manner (Velocity dispersion = Mach number $\times$ Sound speed).  \\\\
The initial linemass, $f_{cyl}$ = 0.5, for Cases 1 to 4. Realisation 5 is a repetition of Case 2, but now with the assumption of isothermality. Finally, to test the hypothesis that the external pressure indeed determines the morphology of filament evolution - irrespective of the initial linemass, realisations 6 and 7 were developed for $P_{ext}/k_{B}\sim$ 10$^{4}$ K cm$^{-3}$, but with respectively, $f_{cyl}$ = 0.7 (transcritical) and 0.3 (sub critical), i.e., a repetition of Case 1, but with a lower $f_{cyl}$. To test convergence, the entire set of realisations was repeated for three choices of the random number seed used for generating the initial turbulent velocity field. \\\\
The fragmentation length scale defined by Eqn. (1) above at 10 K, the minimum achievable temperature in these realisations, and a density $\sim 10^{7}$ cm$^{-3}$, the typical average density of a prestellar core and the maximum resolvable density in this work, is $\lambda_{frag}\sim 0.024$ pc. The actual fragmentation length scale, however, is likely to be somewhat higher due to the turbulence. In the interest of maintaining relatively low computational costs without compromising numerical resolution, we took the average {\small SPH} length, $h_{avg}\sim 0.5\lambda_{frag}$, which satisfies the Hubber criterion for resolving the fragmentation length scale (Hubber \emph{et al.} 2006), the {\small SPH} equivalent of the Truelove criterion for the adaptive mesh refinement algorithm. In other words, the initial radius, $r_{fil}$, of the filament was represented by 17 smoothing lengths so that turbulence within the filament was reasonably well resolved.\\\\ 
The filament is not perturbed axially in any realisation so that any subsequent fragmentation is due to the growth of random perturbations. The number of particles representing the filament is calculated as,
\begin{equation}
N_{gas} = \Big(\frac{3}{32\pi}\Big)\frac{N_{neibs}Vol_{fil}}{h_{avg}^{3}};
\end{equation}
where $N_{neibs}$ = 50, the number of neighbours for each {\small SPH} particle and $Vol_{fil}$ is the volume of the cylinder. The number of particles representing the gas to be accreted and the {\small ICM} were similarly calculated. Thus the entire computational domain was represented with $\sim 1.5$ M particles, of which $\sim 1.1$ M represented the gas while the rest represented the {\small ICM}. Simulations discussed here are therefore reasonably well resolved to capture filament fragmentation. The {\small ICM} is represented by a separate species of {\small SPH} particles that like the gas particles are mobile, but interact through hydrodynamic forces only. Their temperature, $T_{ext}$, is held constant over the duration of a realisation.
\begin{figure*}
  \vspace{1pc}
  \centering
  \mbox{\subcaptionbox{$\frac{P_{ext}}{k_{B}} = 2\times 10^{4}$ K cm$^{-3}$ (Case 1) \label{}}{\includegraphics[angle=270,width=0.45\textwidth]{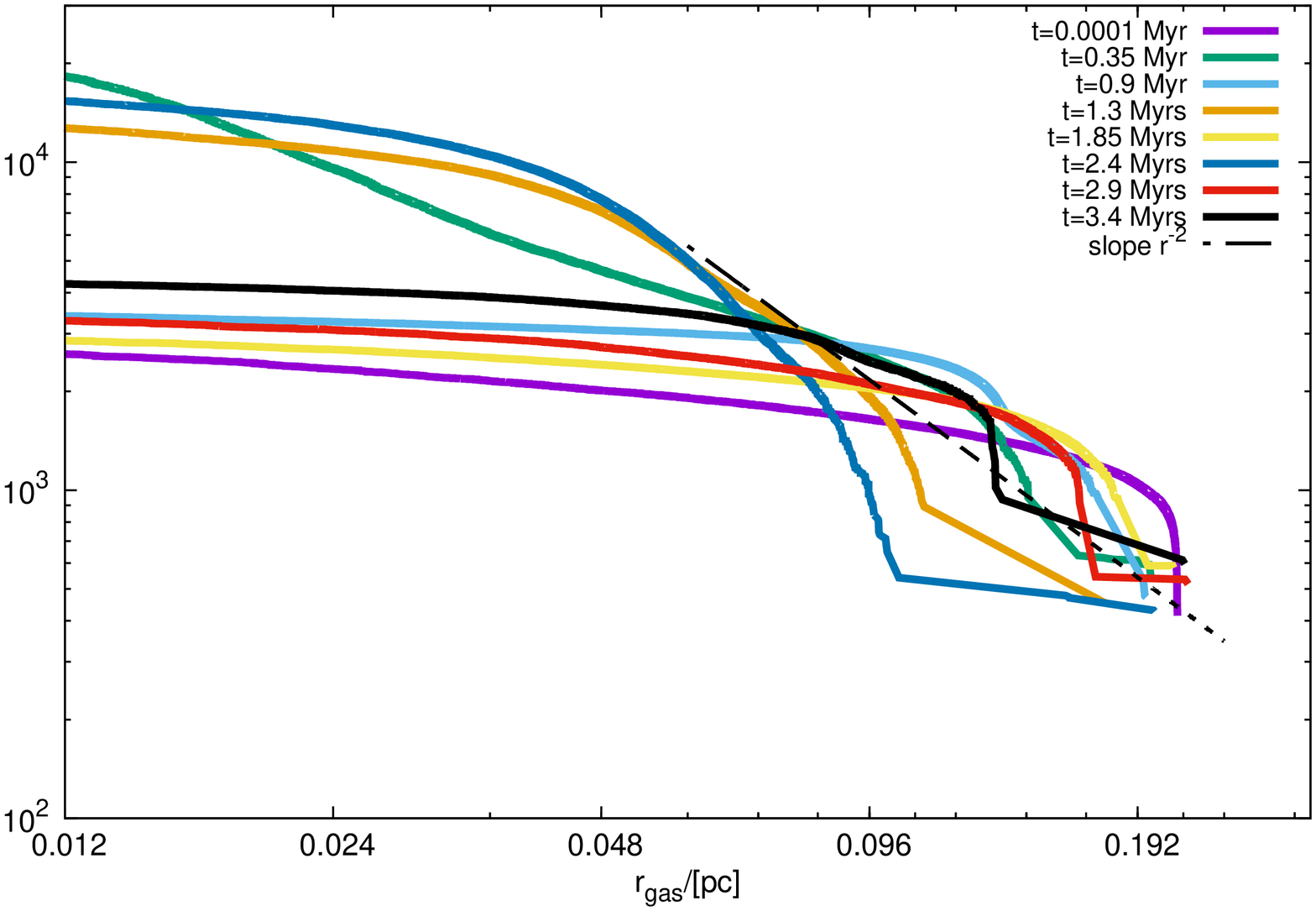}}
        \subcaptionbox{$\frac{P_{ext}}{k_{B}} = 2\times 10^{5}$ K cm$^{-3}$ (Case 2)\label{}}{\includegraphics[angle=270,width=0.45\textwidth]{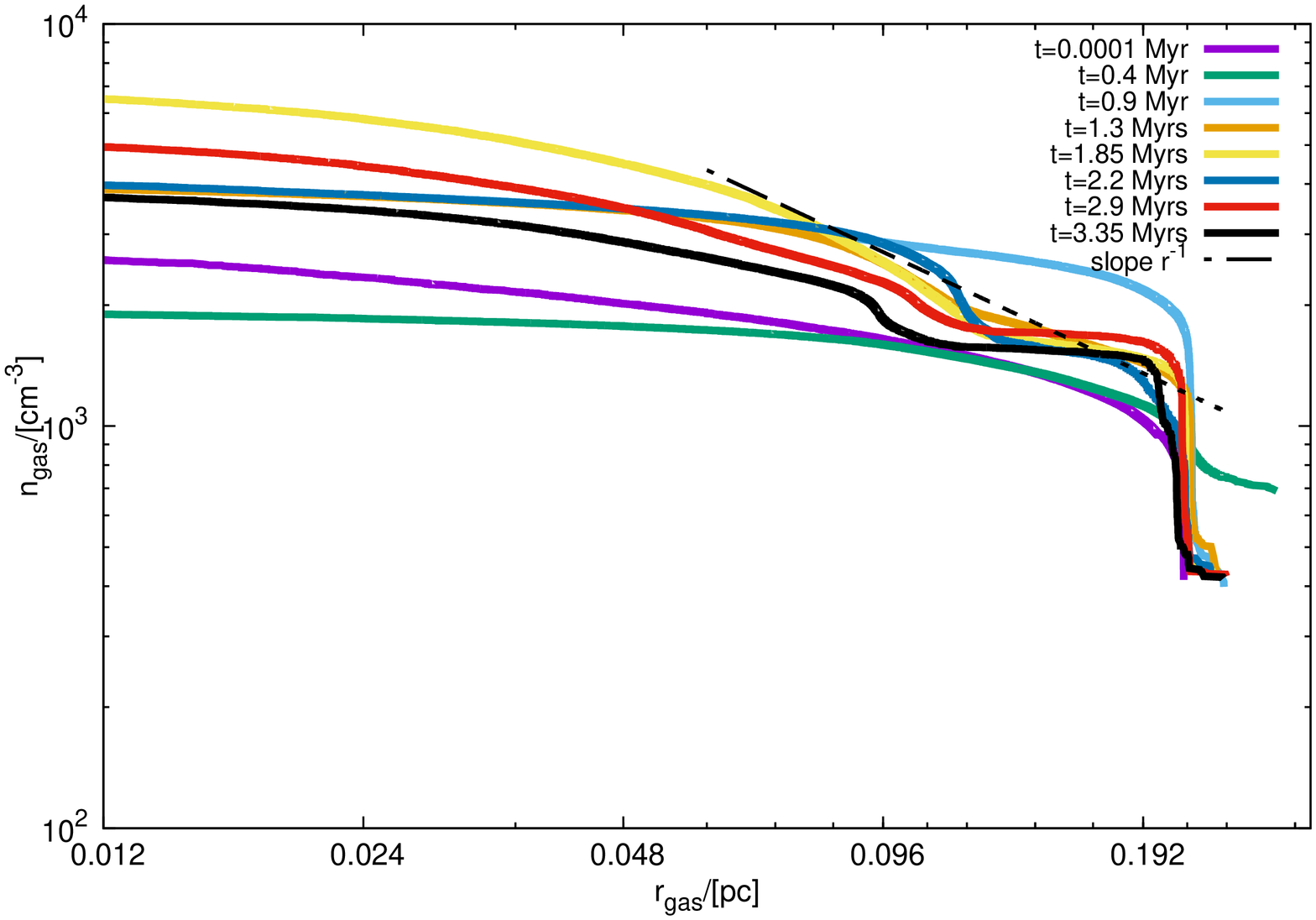}}}
  \mbox{\subcaptionbox{$\frac{P_{ext}}{k_{B}} = 2\times 10^{6}$ K cm$^{-3}$ (Case 3)\label{}}{\includegraphics[angle=270,width=0.45\textwidth]{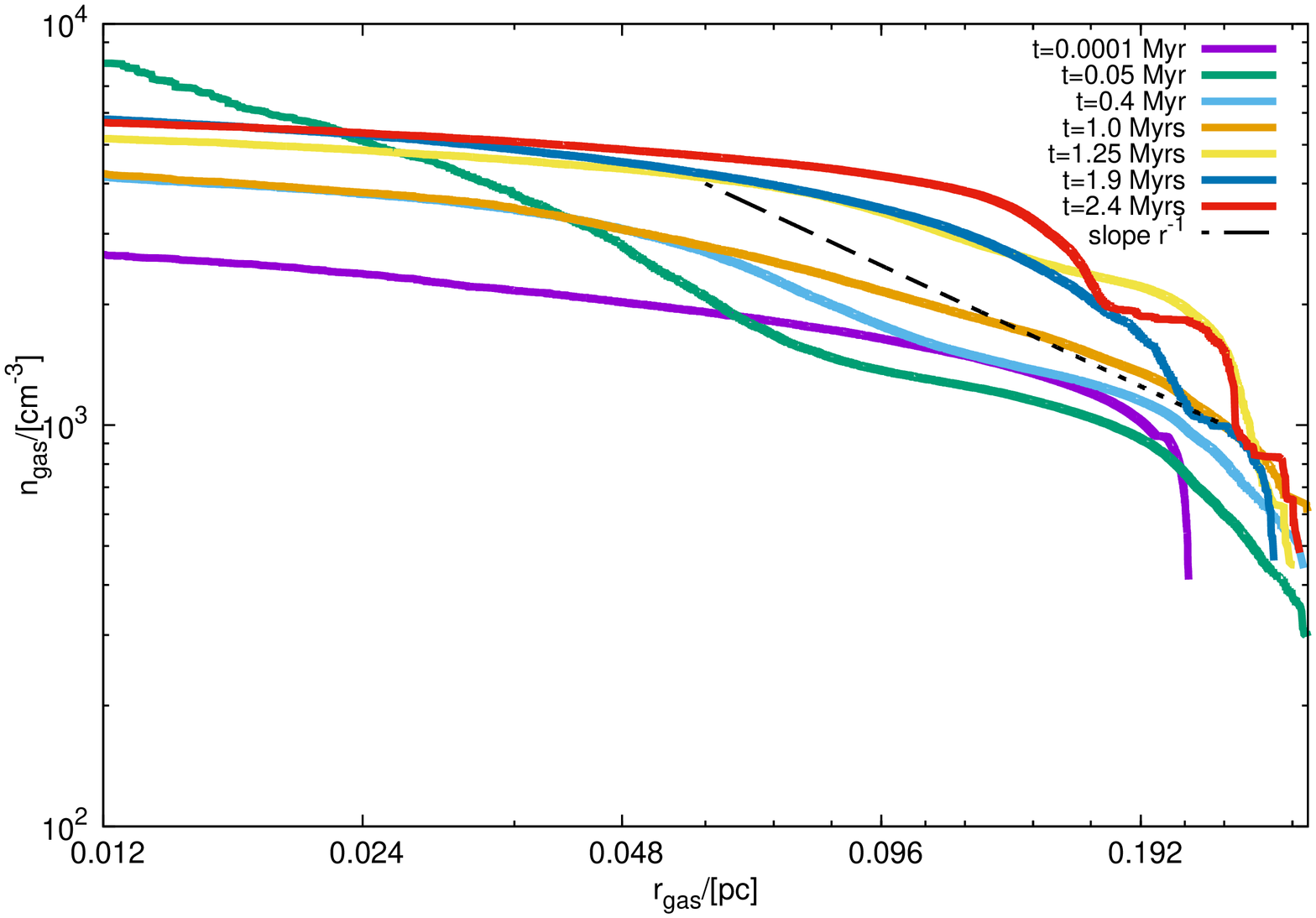}}
        \subcaptionbox{$\frac{P_{ext}}{k_{B}} = 2\times 10^{7}$ K cm$^{-3}$ (Case 4)\label{}}{\includegraphics[angle=270,width=0.45\textwidth]{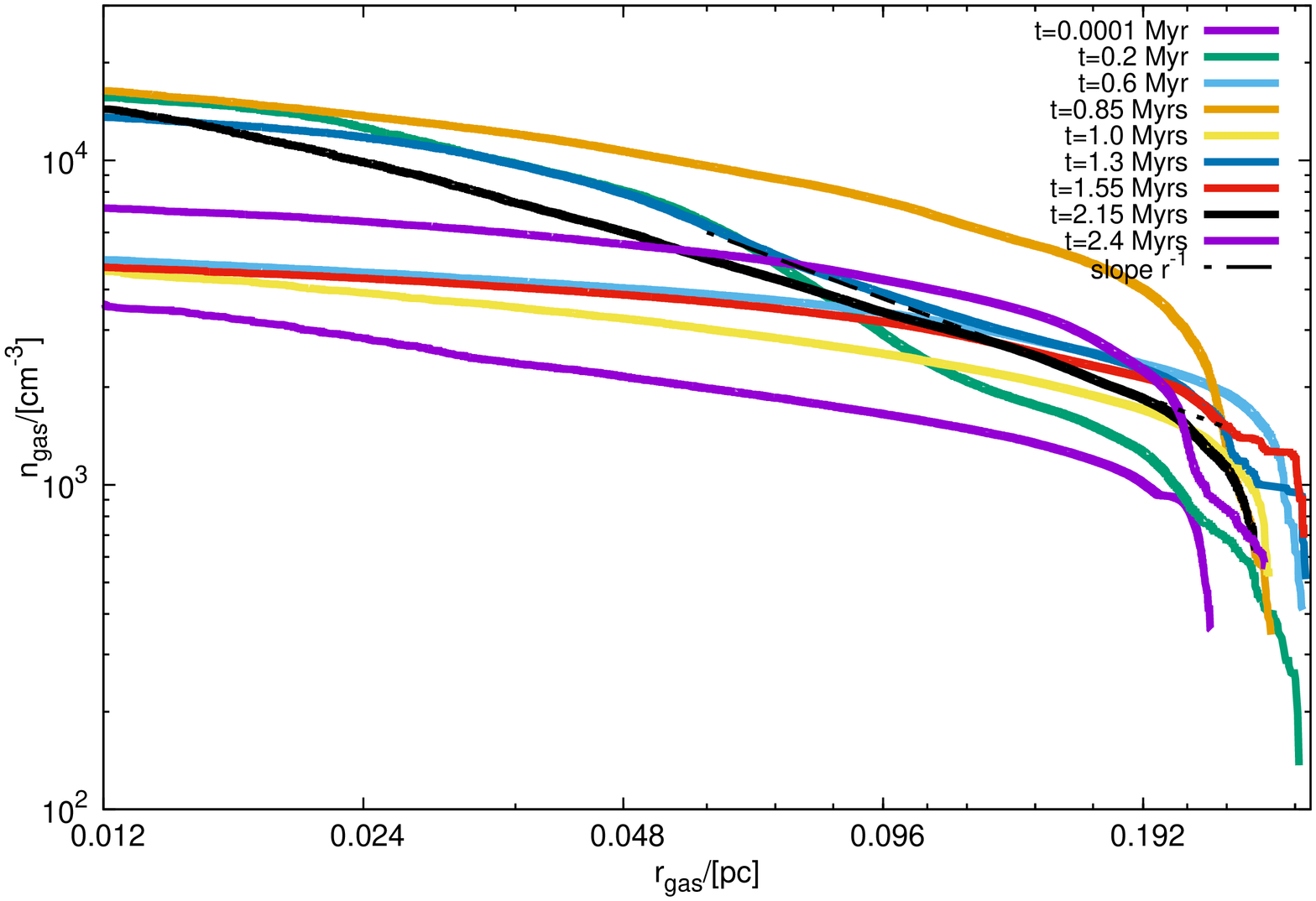}}}
  \mbox{\subcaptionbox{$\frac{P_{ext}}{k_{B}} = 2\times 10^{5}$ K cm$^{-3}$ (Case 5 - isothermal)\label{}}{\includegraphics[angle=270,width=0.45\textwidth]{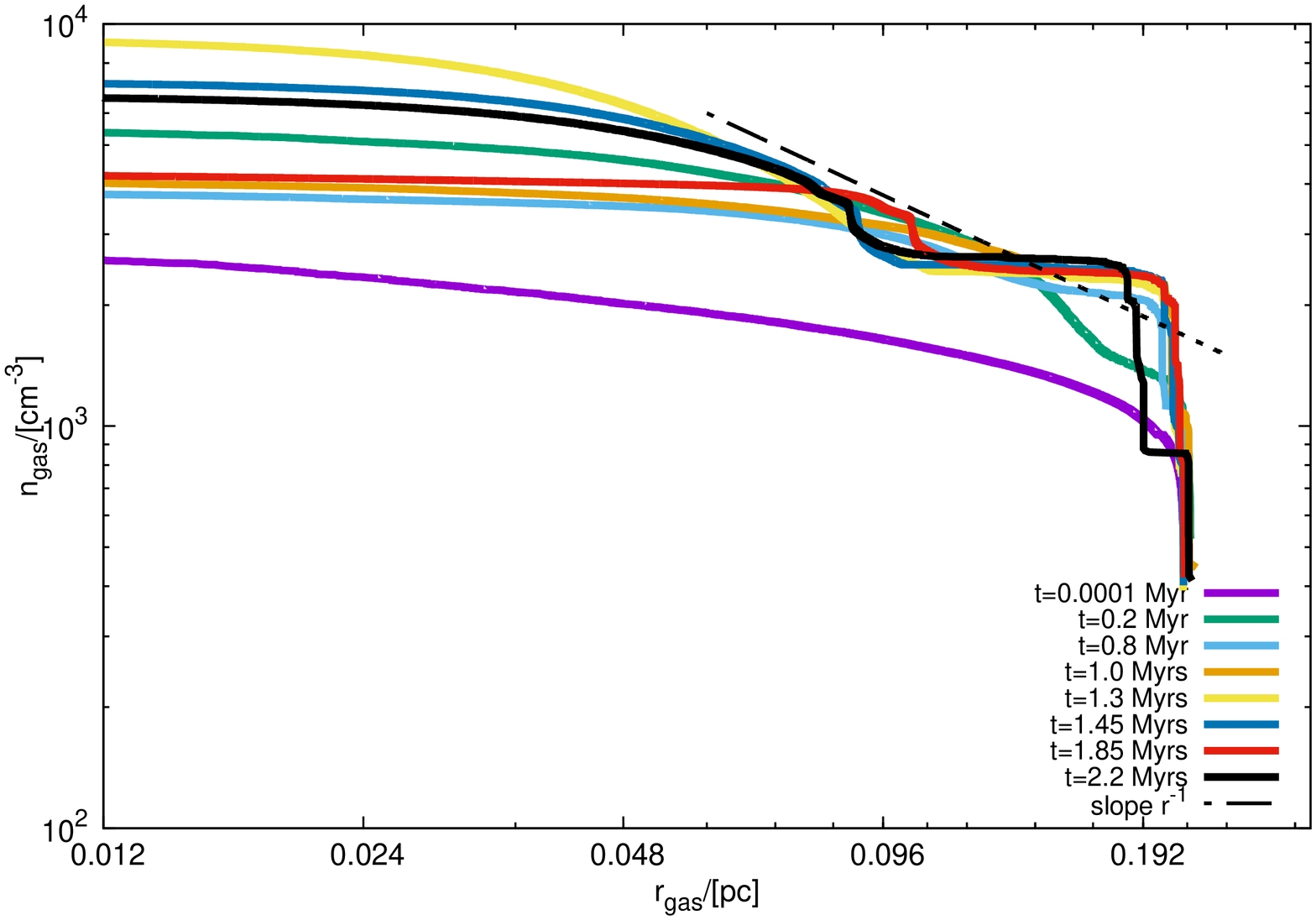}}
        \subcaptionbox{Epoch when a prominent filament spine appeared in respective Cases \label{}}{\includegraphics[angle=270,width=0.45\textwidth]{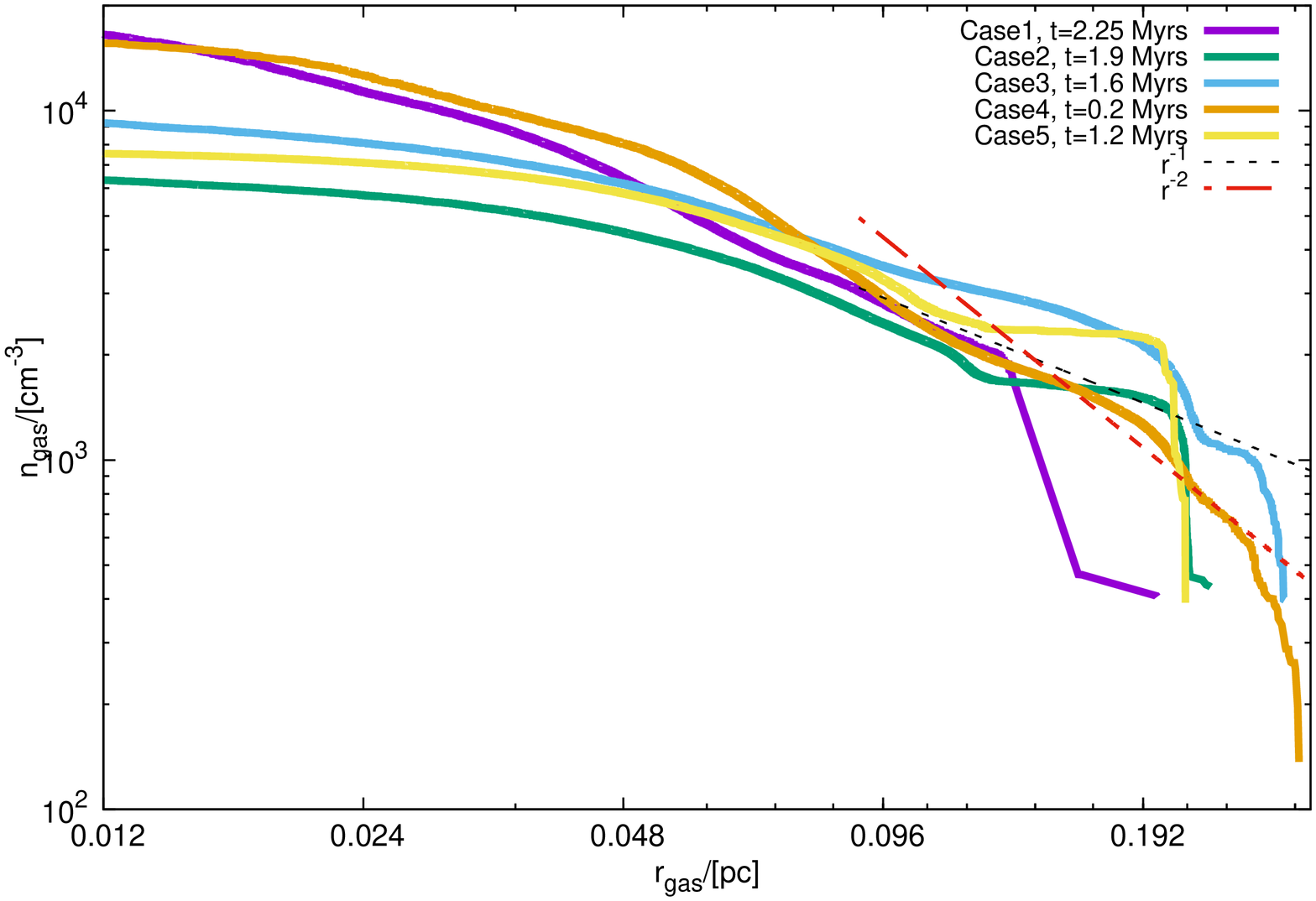}}}
    \caption{Time variation of the radial density profile of the test cylinder in each realisation ($f_{cyl}$ = 0.5). The abscissa $r_{gas}$ is the radius of the filament at a given epoch of time. }
\end{figure*}
\section{Results}
\subsection{Evolution of the test cylinder ($f_{cyl}$ = 0.5)}
The propensity of an isothermal cylinder to contract radially irrespective of whether its linemass, $M_{l}$, is comparable to or exceeds its critical linemass, $M_{l_{crit}}$, is well known. Analytic calculations show that an initially super-critical filament contracts rapidly, rather than collapses, on a time scale much shorter than the growth time scale of perturbations along its length so that the filament in this instance does not fragment along its length (Inutsuka \& Miyama 1992). The test cylinder in each realisation in this set, being initially sub-critical, merely contracted in the radial direction while continuously accreting gas from the external medium, without ever leading to a runaway collapse. The cylinder acquired a centrally peaked density profile following which it then quickly rebounded, giving way to expansion in the radial direction even as it continued to accrete from the surrounding medium.\\\\ 
\begin{figure}
  \vspace{1pc}
  \centering
  \includegraphics[angle=270,width=0.48\textwidth]{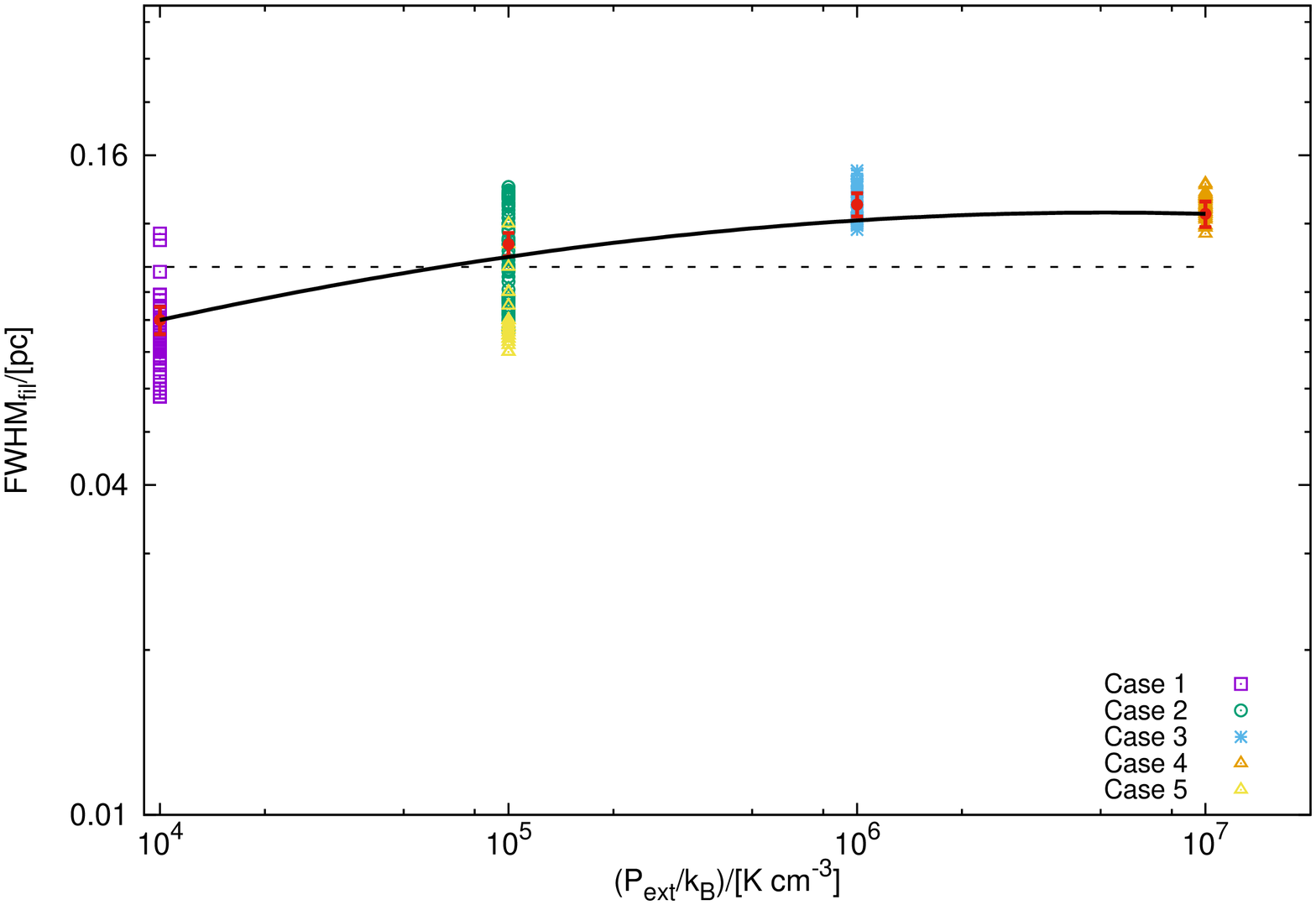}
  \includegraphics[angle=270,width=0.48\textwidth]{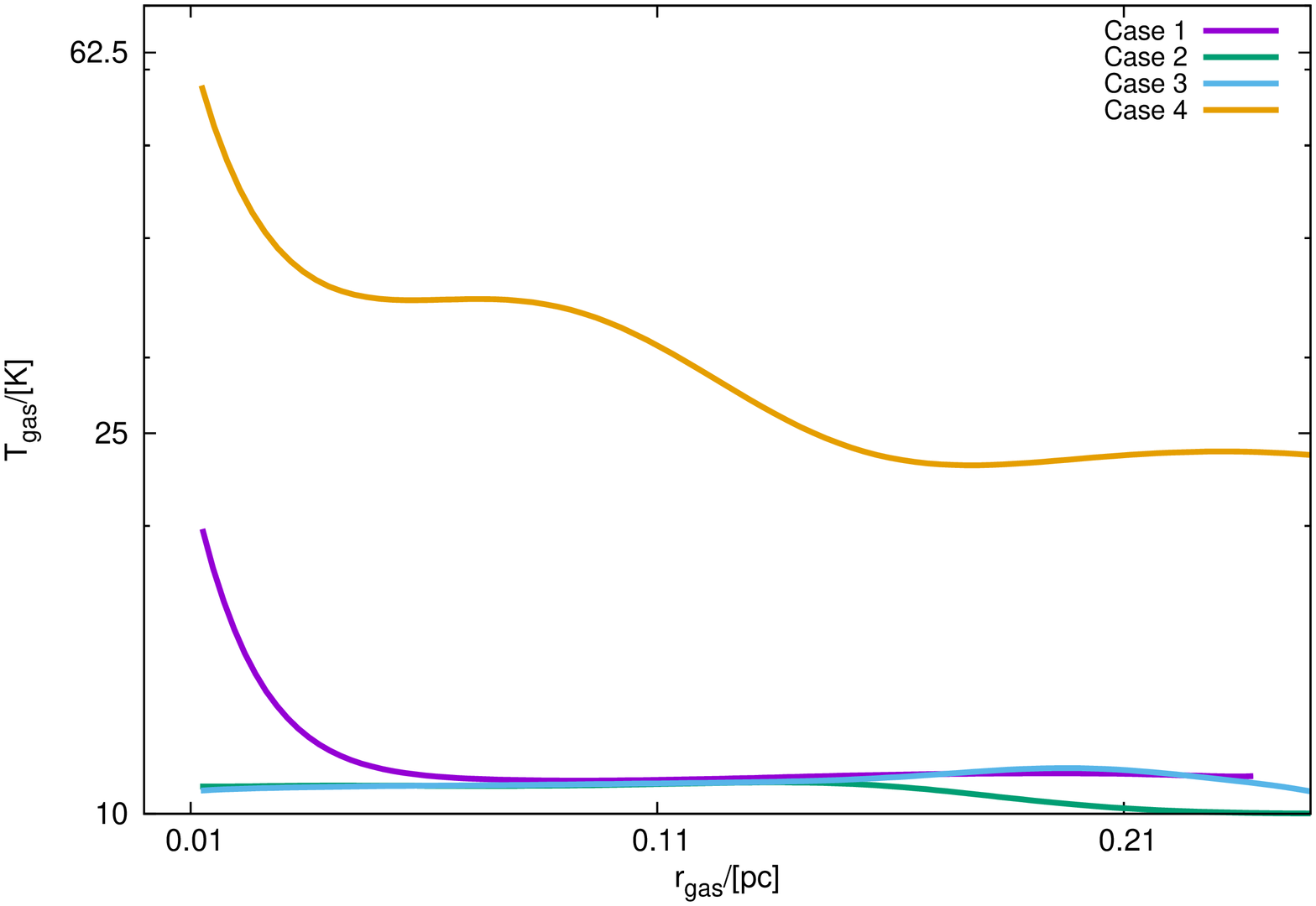}
  \includegraphics[angle=270,width=0.48\textwidth]{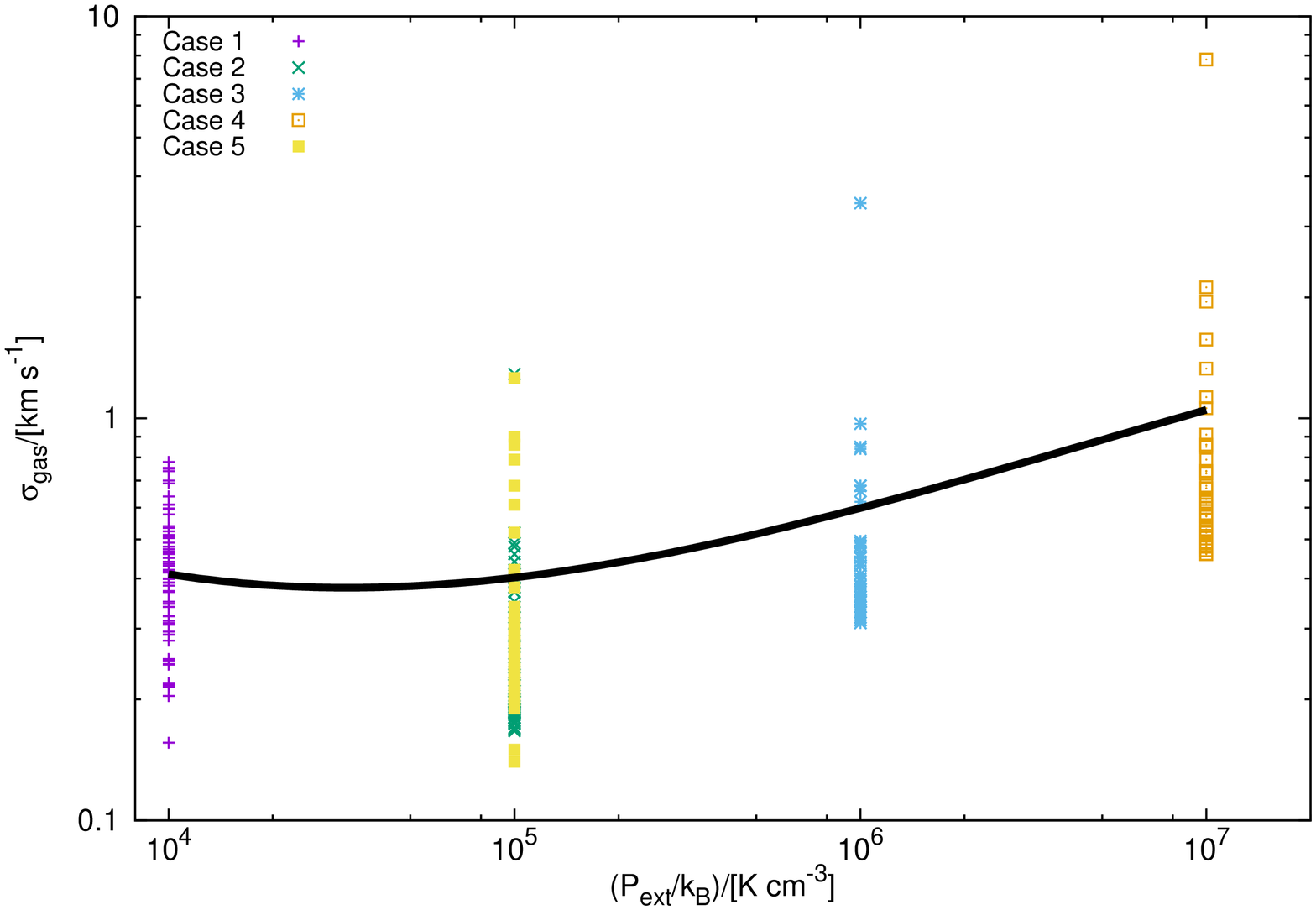}
  \caption{\emph{Top panel :} The $\mathrm{{\small FWHM}_{fil}}$ of the cylinder in each case at different epochs of its evolution is shown on the upper-panel. Dashed line represents $\mathrm{{\small FWHM}_{fil}}$ = 0.1 pc. Red errorbars on this plot represent the mean error in estimation of the $\mathrm{{\small FWHM}_{fil}}$. \emph{Middle panel:} The radial temperature profile of the filament in each case at the epoch when the calculations were terminated. Note that the temperature profile for Case 5 has not been shown here as it is isothermal. \emph{Bottom panel :} The mean velocity dispersion, $\sigma_{gas}$, in the filament in each case at different epochs of its evolution. }
\end{figure}
\begin{figure}
  \vspace{1pc}
  \centering
  \includegraphics[angle=270,width=0.48\textwidth]{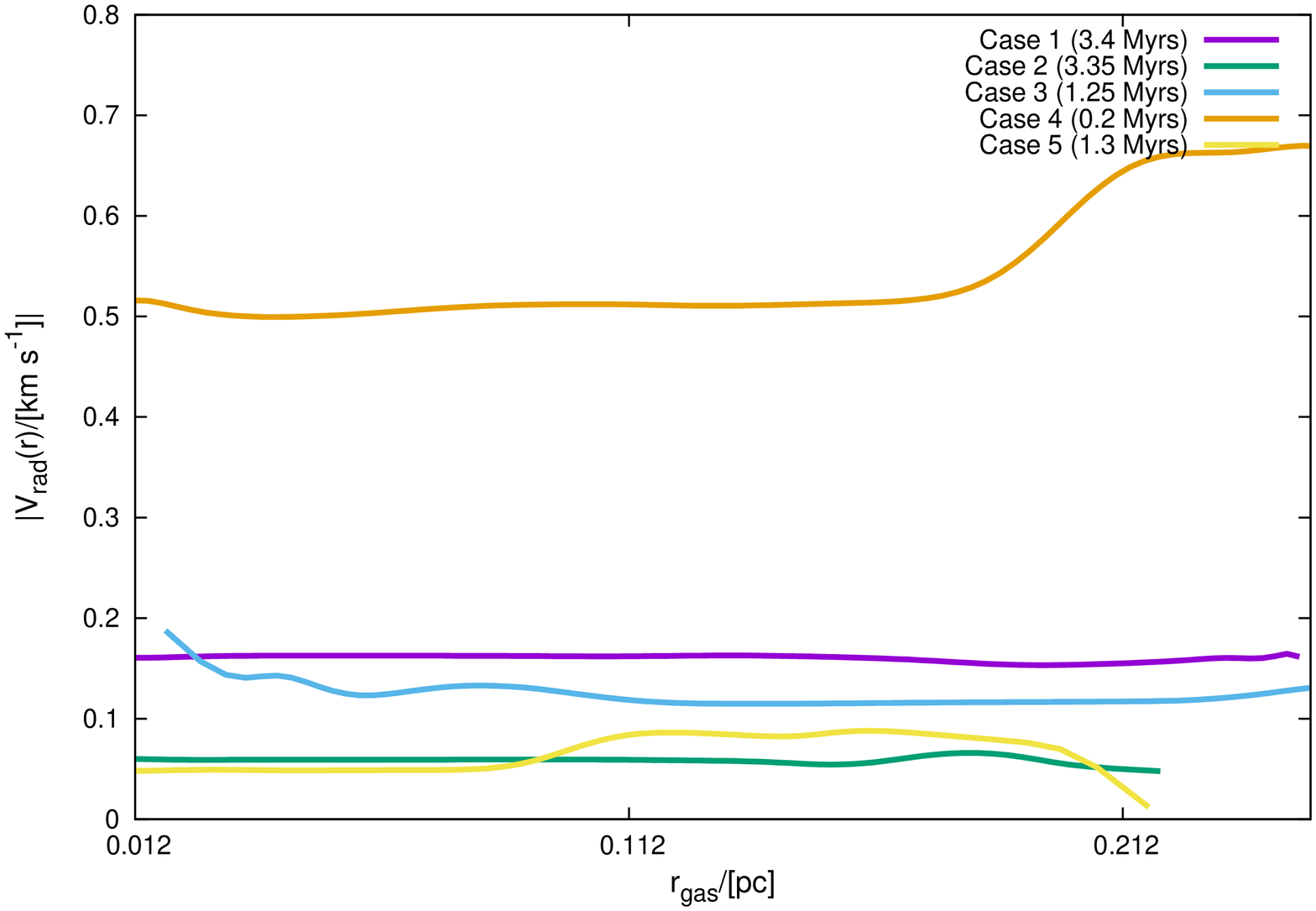}
  \caption{Magnitude of density weighted radial velocity at the epoch when the filament in each realisation acquired its peak central density.}
\end{figure}
The accreting test cylinder, i.e., filament soon stabilised and thereafter exhibited only a relatively small variation in its peak density. Panels (a) to (e) in Fig. 2 show plots of the radial density profile of the filament at different epochs of its evolution in each realisation.
Irrespective of the external pressure due to the {\small ICM}, the respective density profiles are qualitatively similar, modulo a slight variation in the peak density. The slope of the density profile in the outer regions of the filament varies between $n_{gas}\propto r^{-1}$ and $n_{gas}\propto r^{-2}$ across the realisations developed in this work. In fact, it tends to become shallower for external pressure $\ge 10^{5}$ K cm$^{-3}$. For example, panels (b), (c) and (d) of Figure 2 show density profiles where the slope in outer regions of the filament appears consistent with $r^{-1}$, while in Case 1 the slope varies as $r^{-2}$. \\\\ 
The slopes of the radial density profiles observed here are roughly consistent with those often reported for typical filaments in nearby star-forming {\small MCs} (e.g., Lada \emph{et al.} 1999; Arzoumanian \emph{et al.} 2011), who reported density profiles falling off as $r^{-1.5} $ to $r^{-2.5}$, but are significantly shallower than the slope in the Ostriker (1964) model for an isothermal filament ($n_{gas}\propto r^{-4}$). Finally, panel (f) of Fig. 2 shows the density profile of the filament at the epoch when it exhibited a prominent spine (i.e., the \emph{inner} filament). We note that this epoch is not the one where the filament has acquired its first central density peak due to radial contraction. For as noted above, the filament rebounds after this initial contraction. Instead, the density profiles shown correspond to the epoch when the filament in respective cases readily exhibits a spine in the rendered density images; see \S 3.3 below. As observed at other epochs of its evolution, the slope of the density profile in the outer regions of the filament even at this epoch is shallow, having a slope between $r^{-2}$ and $r^{-1}$. \emph{Evidently, external pressure has some influence on the density profile of a filament.} Below we explore whether or not the magnitude of external pressure also affects other filament properties, including their fragmentation. 
\subsection{Physical properties of the filaments}
The upper-panel of Fig. 3 shows the filament width, i.e., its full width at half maximum ($\mathrm{{\small FWHM}_{fil}}$), for different choices of external pressure, $P_{ext}$. Multiple points for a given magnitude of external pressure correspond to the filament width at different epochs of its evolution. The width of the lognormal distribution fitted across the filament was measured at various locations along its length with $\sigma_{r}$ being the dispersion of the corresponding estimates of the width. Finally,  $\mathrm{{\small FWHM}_{fil}} = 2(\sqrt{2\ln 2})\sigma_{r}$. A sample calculation of the $\mathrm{\small FWHM}_{fil}$ is shown in Appendix A below. The continuous solid black line in this plot corresponds to the median $\mathrm{{\small FWHM}_{fil}}$ for each realisation. The tendency for slight broadening of the filament at higher magnitudes of external pressure is readily visible from this plot. \\\\
Seeing larger widths at larger external pressures is counter-intuitive and inconsistent with the conclusions of Fischera \& Martin (2012) who predicted the filament width varies with external pressure as $\mathrm{{\small FWHM}_{fil}}\propto P_{ext}^{-0.5}$. We note that the filament in the isothermal realisation (Case 5) has the smallest width which could perhaps be a consequence of stronger self-gravity, a point we will investigate later in \S 3.3.1. Observations of filaments in nearby {\small MCs}, however, show that they have approximately uniform width on the order of $\sim$0.1 pc (Arzoumanian \emph{et al.} 2011, 2019). Our findings here that suggest a slight increase in filament width with increasing external pressure contrasts with that of a universal filament width by Arzoumanian et al. At the most, we can say that the filament width is on the order of $\sim$0.1 pc for external pressures comparable to that in the Solar neighbourhood. These findings motivate one to inquire into the factors that stabilise a filament against self-gravity. We discuss this point further in \S 4 below.\\\\
In the absence of a magnetic field, thermal pressure and pressure due to turbulence injected by accreting gas are the two agents that must support the filament against self-gravity. This situation is clear from the remaining two plots in Fig. 3. First, the central panel shows the terminal temperature profiles of the model filaments. The filament in each case acquires an approximately isothermal configuration by the time the realisation is terminated, but gas within the filament remains relatively warm in Case 4 ($P_{ext}/k_{B}\sim 10^{7}$ K cm$^{-3}$). Evidently, given the observed filament widths, which are an order of magnitude greater than the local thermal Jeans length, thermal support alone cannot be sufficient to support a filament against self-gravity. Second, data points in the lower panel of Fig. 3 represent the magnitude of the mean velocity dispersion ($\sigma_{gas}$) within the filament at different epochs of its evolution. As before, the solid black line on this plot represents the median velocity dispersion for each realisation. \\\\
Observe that the mean velocity dispersion is the smallest in the first realisation where the gas within the filament was  initially subsonic and thereafter it increases steadily with increasing external pressure. Together these plots demonstrate the importance of injected turbulence over thermal energy in supporting a filament against self-gravity. Greater buoyancy due to a stronger induced turbulence at a greater magnitude of external pressure possibly explains the relative shallowness of the respective density profiles observed in Fig. 2. This observation of the relative shallowness of the density profile also holds in the isothermal realisation of Case 5. \emph{The $\mathrm{{\small FWHM}_{fil}}$ of a filament as seen on the upper-panel of Fig. 3 depends only weakly on the magnitude of external pressure or equivalently, on the velocity of the gas being accreted}.\\\\
Figure 4 shows the magnitude of the density weighted radial velocity, $V_{rad}$, for each realisation at the epoch when the filament acquired its peak density. In Case 4, which has the highest external pressure, the filament initially contracts rapidly, as reflected by it having the largest $V_{rad}$. The corresponding density profile at this epoch, as can be seen in Fig. 2 ($t\sim 0.2$ Myrs, green line in the corresponding plot), is also relatively steep with $n_{gas}\propto r^{-2}$. In comparison, the interiors of the filament do not exhibit a steep velocity gradient in any realisation, suggesting that, far from a runaway collapse, the filaments in these realisations simply contract in a quasistatic manner via the interplay between self-gravity, thermal pressure and pressure due to turbulence within the filament.   
\subsection{Morphology of filament fragmentation}
The filament in each realisation begins to fragment even as it contracts. Given that the model filaments here are cylinders with finite dimensions, accumulation of gas at its ends is observed and locally collapsing regions first appear there. Pile-up of gas at the edges, sometimes also referred to as end dominated collapse (Bastien 1983), is often observed in clouds having idealised geometry (e.g., Burkert \& Hartmann 2004). Rendered density images on various panels of Fig. 5(a) show the appearance of unstable modes along the filament at different epochs of its evolution. A perusal of these panels shows that the filament appears susceptible to the \emph{sausage type} instability, at least during the early stages of its evolution irrespective of the magnitude of external pressure. The end product of this evolutionary sequence, however, is remarkably different for a filament experiencing a relatively higher external pressure ($\gtrsim 10^{6}$ K cm$^{-3}$). This behaviour is also true for the isothermal realisation in Case 5, as can be seen in Fig. 5(b). Equivalently, the choice of equation of state does not have much bearing on the morphology of filament fragmentation. For external pressure $\ge 10^{6}$ K cm$^{-3}$, i.e., in Cases 3 and 4, however, the filament can additionally be seen to buckle. At a relatively low external pressure, as in Case 1 ($P_{ext}/k_{B}\sim 10^{4}$ K cm$^{-3}$), the fragments are \emph{pinched}. By contrast, in the remaining realisations, however, the fragments are \emph{broad}. The terms \emph{pinched} and \emph{broad} are used in the same connotation here as by, for instance Heigl \emph{et al.} (2019). \\\\
\begin{figure*}
\begin{subfigure}{160mm}
  \vspace*{1pt}
  \includegraphics[angle=0,width=\textwidth]{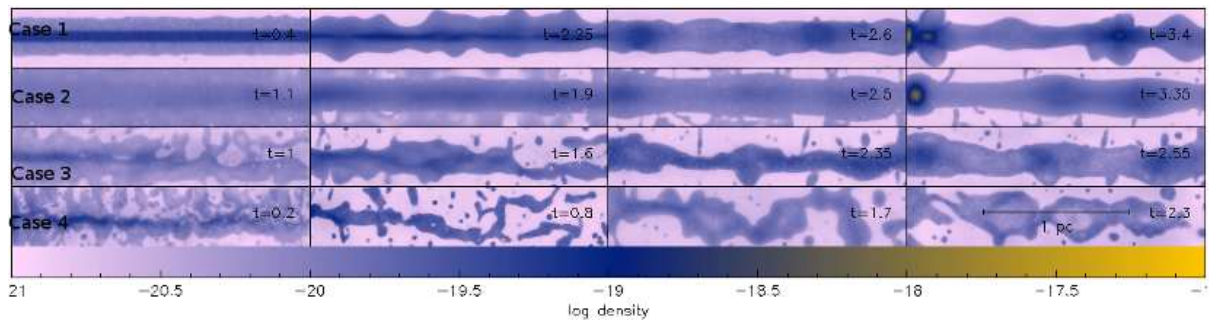}
  \caption{Time in Myrs is marked at the top right-hand corner of each panel.}
\end{subfigure}
\begin{subfigure}{160mm}
  \vspace*{1pt}
  \includegraphics[angle=0,width=\textwidth]{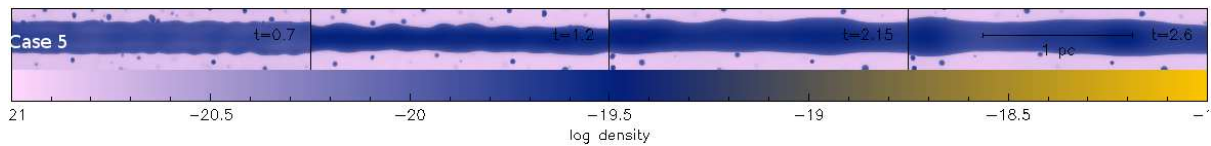}
  \caption{As in (a), but now for the isothermal realisation Case 5.}
\end{subfigure}
\caption{Rendered density images of a cross-section in the plane of the filament in each case at different epochs of its evolution.}
\end{figure*}
The zoomed-in images in Fig. 6(a) illustrate the morphology of fragments more clearly. The circular, stratified density profile similar to that of a stable Bonnor-Ebert sphere, for fragments in the filament confined by an external pressure $\ge 10^{5}$ K cm$^{-3}$, is evident from these plots. Indeed, the density profiles in the latter instances are significantly different from that of the fragment in Case 1 where the external pressure was on the order of $\sim 10^{4}$ K cm$^{-3}$. The tendency to form broad, spherical fragments at relatively high external pressure was also reported by Inutsuka \& Miyama (1997). In some cases, they also observed merger of fragments, as is indeed visible in the panel corresponding to $t$ = 3.35 Myr for Case 2 in Fig. 5(a). Inutsuka \& Miyama (1997) argued that fragments merge when the wavenumber corresponding to this large-scale motion of fragments is half the wavenumber of the fastest growing mode. We note, however, that the findings reported by these authors were in fact in relation to a transcritical filament ($f_{cyl}$ = 1), unlike the case here. \\\\
In the relatively simple classical picture of Inutsuka \& Miyama (1992), fragmentation of a filament depends solely on its linemass, but our realisations here show that the ambient environment is also crucial towards determining the manner in which a filament fragments. This result is also consistent with the findings of Fischera \& Martin (2012). We will briefly examine below the impact of variation of the linemass on filament evolution. Finally, we observe that a filament at significantly large external pressure is likely to rupture as is indeed observed in Case 4 ($P_{ext}/k_{B } = 2\times 10^{7}$ K cm$^{-3}$). Realisations in Cases 1 to 4 in Table 1 were repeated for three choices of the initial seed for the random number generator. In the interest of brevity, however, we show here only the terminal epoch of the filament in Case 2 in Fig. 6(b). Pictures in this Figure demonstrate convergence in the wake of variation in the choice of the initial seed for the random number generator.\\\\    
\begin{figure*}
\begin{subfigure}{160mm}
  \vspace*{1pt}
  \mbox{\includegraphics[angle=0,width=0.33\textwidth,height=30mm]{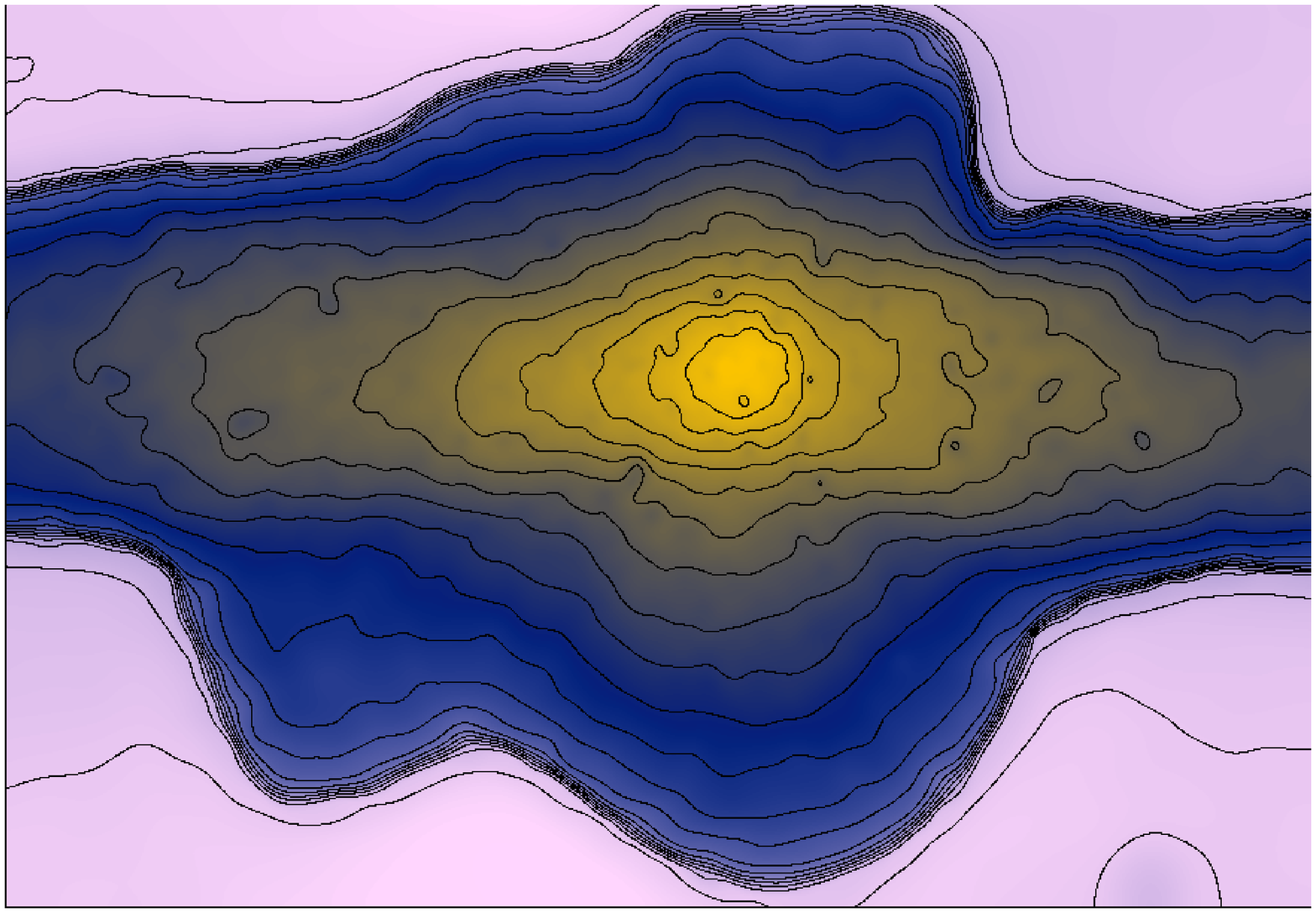}
        \includegraphics[angle=0,width=0.33\textwidth,height=30mm]{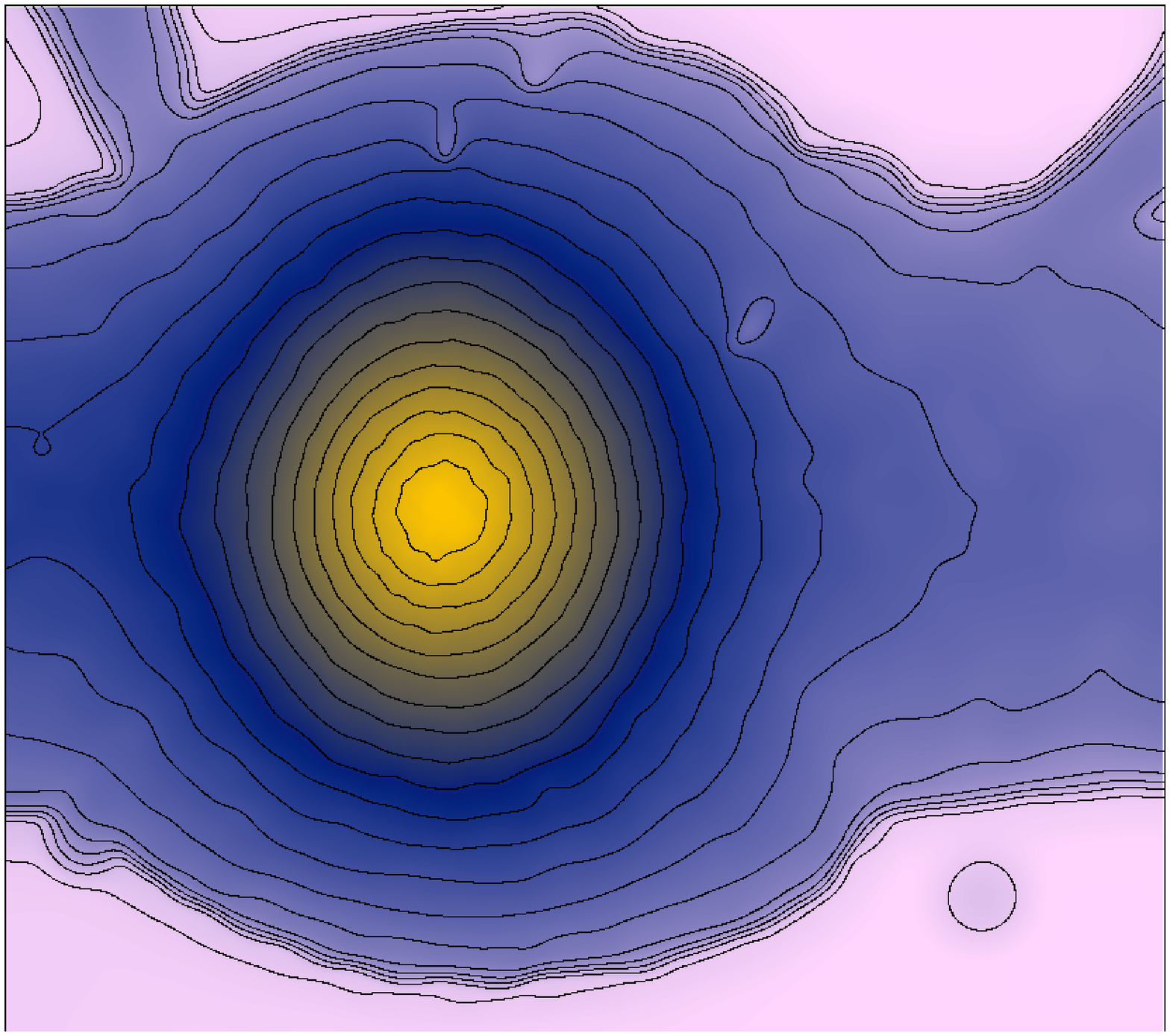}
        \includegraphics[angle=0,width=0.33\textwidth,height=30mm]{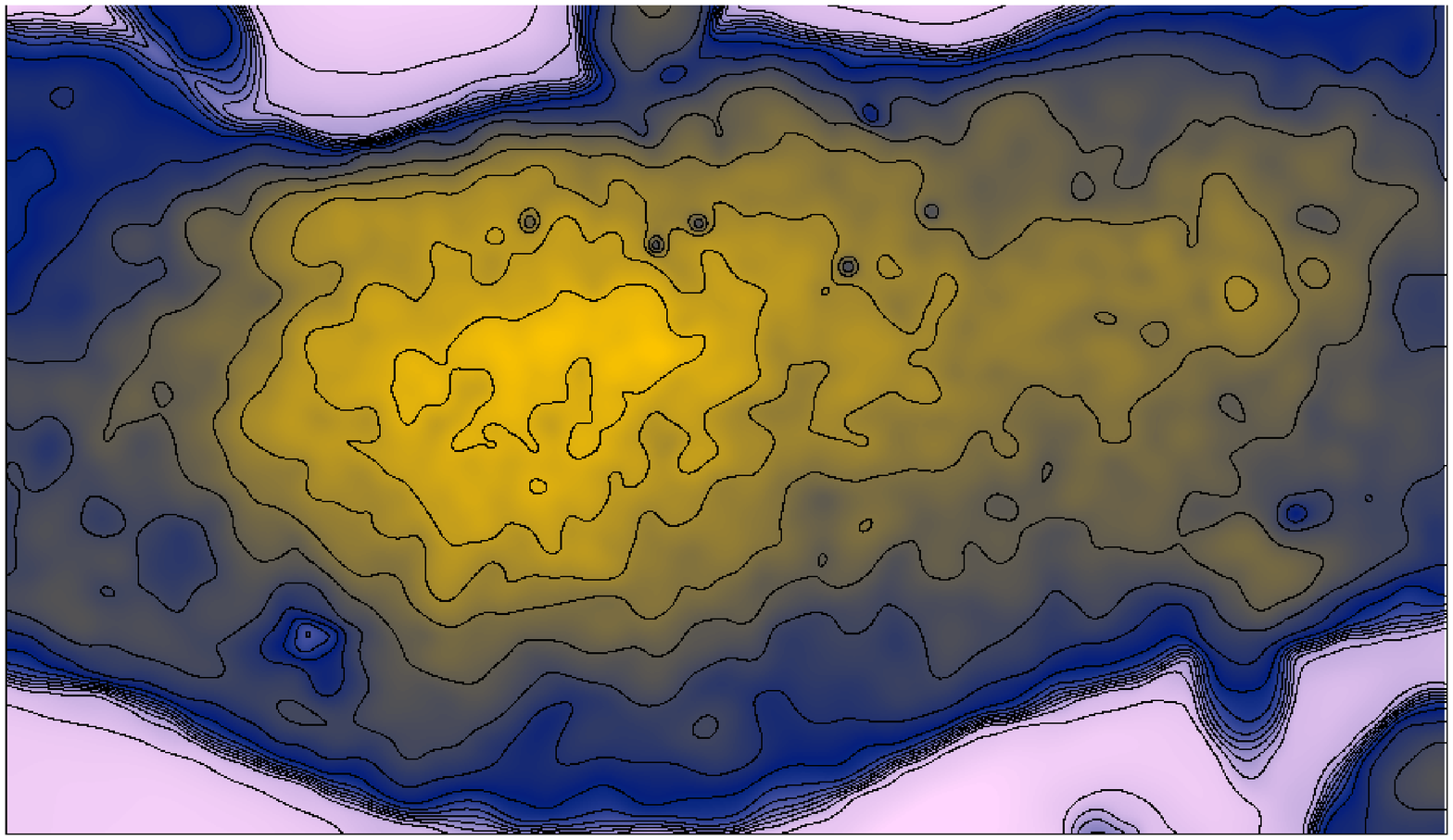}}
  \caption{From left to right, images showing the cross-section of a fragment at the terminal epoch in Cases 1, 2 and 3, respectively. Uniformly placed density contours have been overlaid on these images.}
\end{subfigure}
\begin{subfigure}{160mm}
  \vspace*{1pt}
        \includegraphics[angle=0,width=\textwidth]{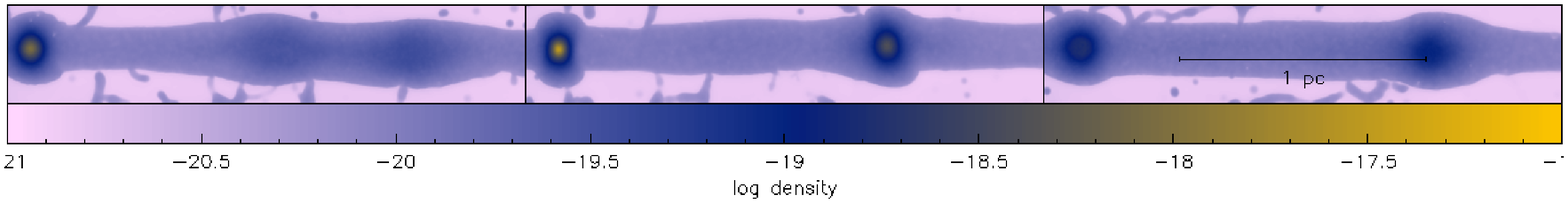}
  \caption{Terminal epoch of the filament in Case 2 for three choices of the initial seed for the random number generator.}
\end{subfigure}
\caption{Rendered density images of a cross-section in the plane of the filament as in Fig. 5.}
\end{figure*}
 The susceptibility of a filament to the \emph{sausage instability} or the \emph{compressional instability} is determined by the magnitude of the ratio, $r_{in}/H$, where $r_{in}$ is the inner radius of the filament and, $H$, the scale height defined by Eqn. (2) above. Nagasawa (1987) showed analytically that a filament with $\frac{r_{in}}{H}\lesssim$ 2 is susceptible to the \emph{sausage-like} deformation instability that manifests by the growth of perturbations on its surface accompanied with fluctuations in pressure along these deformations. On the other hand, a filament with $\frac{r_{in}}{H}\gtrsim$ 2 is susceptible to a Jeans-like \emph{compressional} instability that manifests in the form of local density enhancements which leads to an increase in the self-gravity. Fig. 7 shows the magnitude of this ratio as the filament in each realisation evolved. Note that this ratio is well below 2 in each case. Evidently the filament in each case fragmented via the growth of the \emph{sausage-type} instability as is visible from the rendered images on panels of Fig. 5 (a) and (b). \\\\
The magnitude of the quantity $(\frac{r_{in}}{H})$ tends to increase considerably towards the latter stages of the realisation in Case 1 (and to some extent in Case 2) meaning that the fragments within the filament in these cases could possibly break further via the Jeans-like \emph{compressional} instability. Of course the subsequent evolution of these fragments could be verified only if the realisations were allowed to run further. Nevertheless, we will investigate this problem in our forthcoming article. Case 1 is also the realisation in which \emph{pinched} fragments were seen; see the picture on left-hand panel of Fig. 6(a). The magnitude of this ratio tends to decrease with increasing external pressure as is clear from the line corresponding to Case 4 that was developed with the highest external pressure in the current ensemble. Although the central density increases on a relatively short timescale at a higher external pressure, the velocity dispersion within the filament is also stronger which renders the filament unstable and causes it buckle, as seen in Fig. 5(a). In fact, the pictures on the lower two panels of Fig. 5(a) tend to support the well-known \emph{fray and fragment} scenario in which the natal filament is composed of a bundle of fibres that may eventually fragment. We note, the ratio $(\frac{r_{in}}{H})$ has presently been calculated for the natal filament and not the individual fibres visible for example, on the right-hand panel of the montage for Case 4 in Fig. 5(a). \\\\
\begin{figure}
 \vspace*{1pt}
 \includegraphics[angle=270,width=0.5\textwidth]{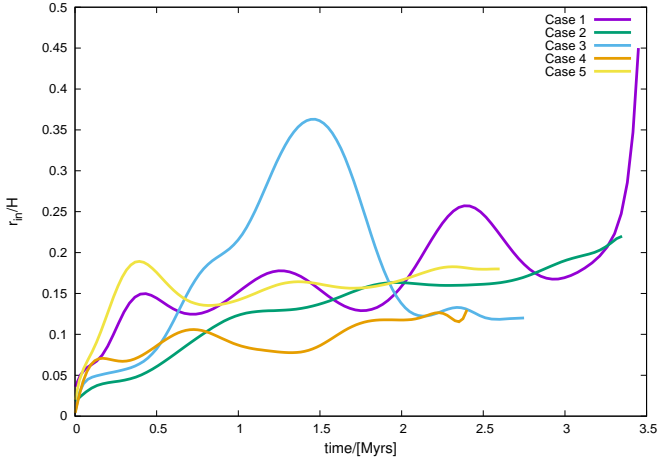}
 \caption{Time variation of the stability factor ($\frac{r_{in}}{H}$) for the filament in each realisation. }
\end{figure}
\begin{figure} [!H]
  \vspace{1pc}
  \centering
  \includegraphics[angle=270,width=0.45\textwidth]{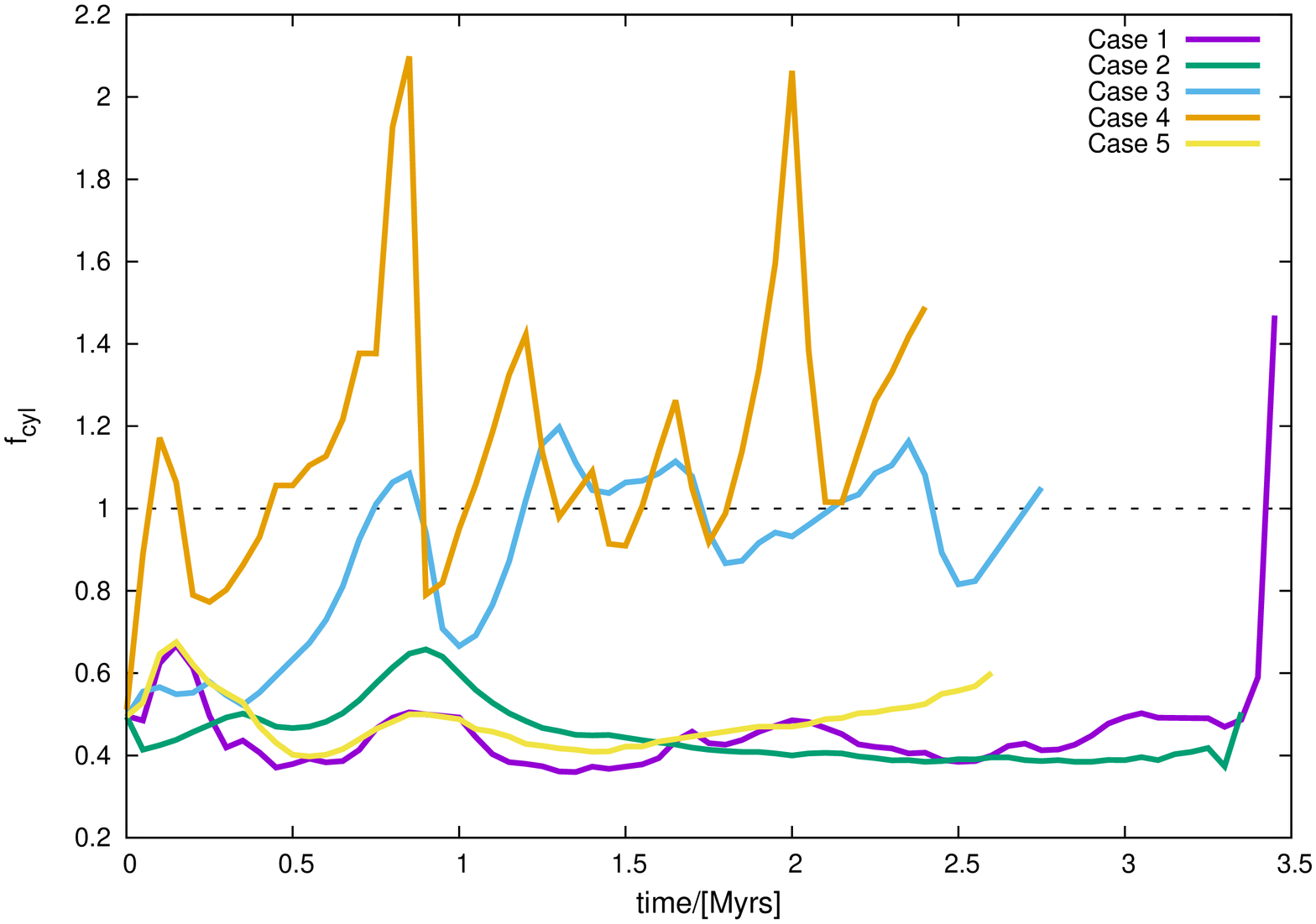}
  \includegraphics[angle=270,width=0.45\textwidth]{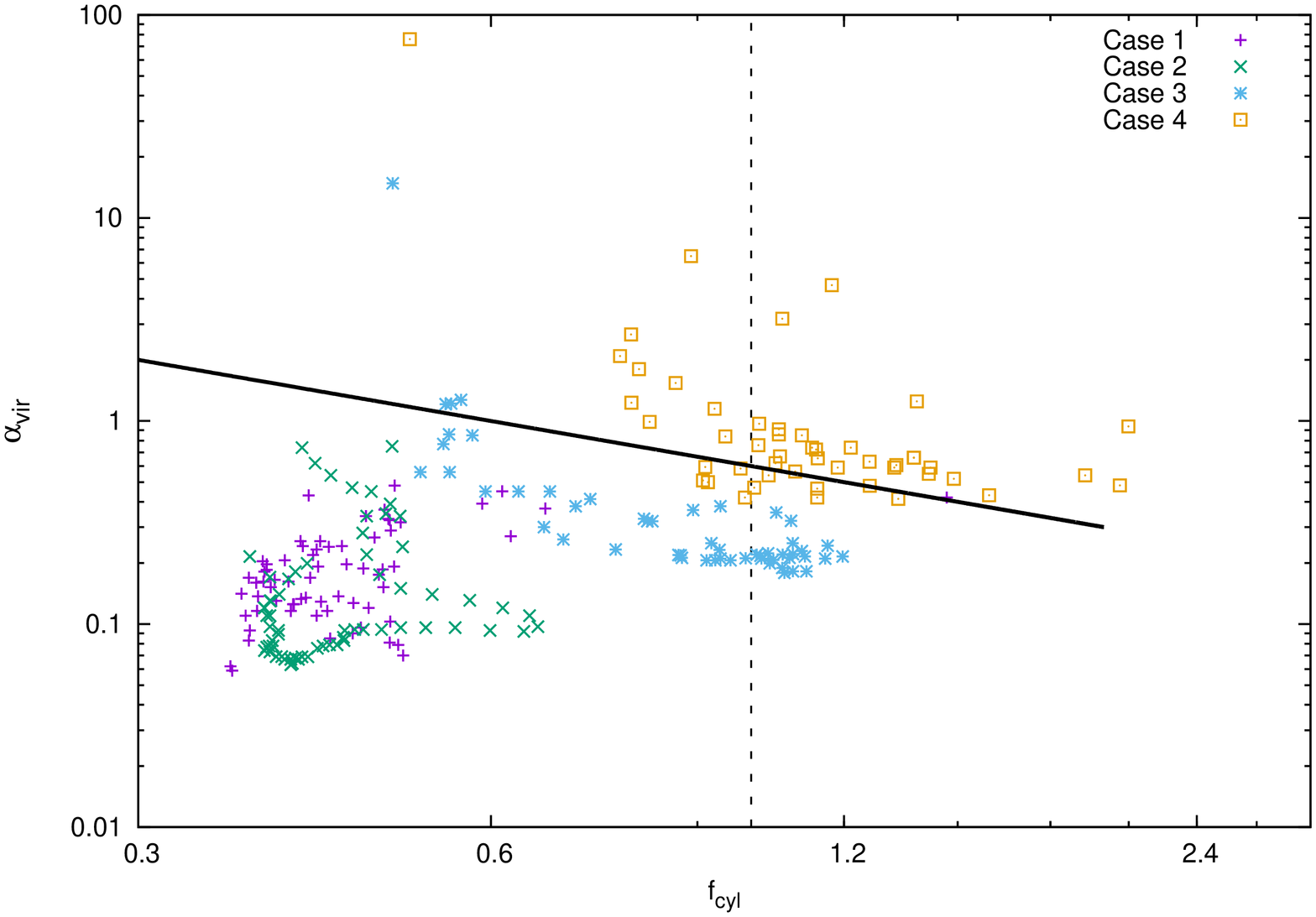}
  \includegraphics[angle=270,width=0.45\textwidth]{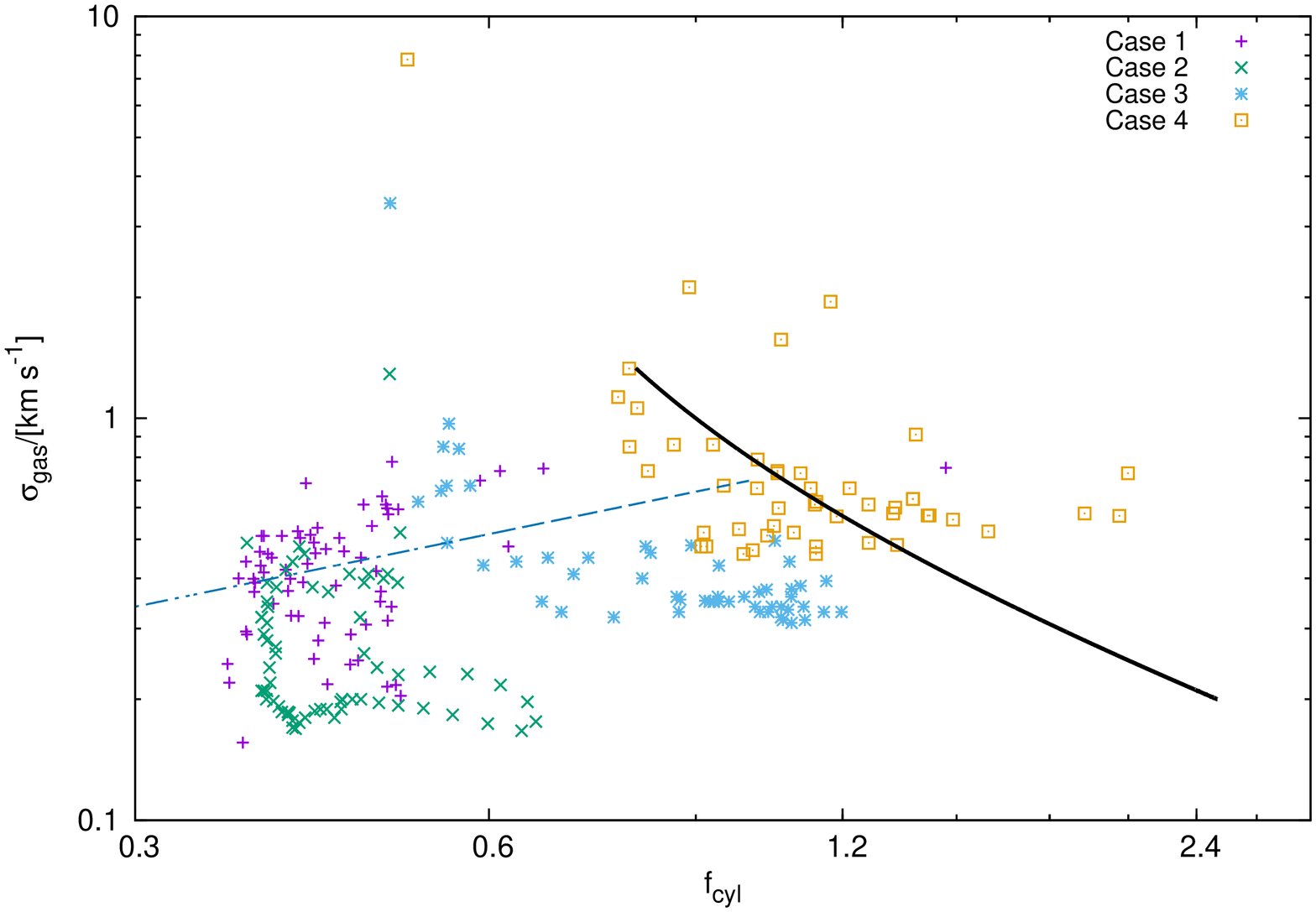}
  \caption{\emph{Top panel :} The time-variation of the linemass. \emph{Middle panel :} The correlation between the Virial parameter, $\alpha_{vir}$, and the linemass. The dashed continuous line in either plot represents the critical linemass, $f_{cyl} = 1$. The bold black line on this plot represents $\alpha_{vir}\propto M_{l}^{-1}$. \emph{Bottom panel :} Correlation between the velocity dispersion, $\sigma_{gas}$, and the filament linemass. The dashed blue line corresponds to $\sigma_{gas}\propto f_{cyl}^{0.5}$. On the contrary, the black line here corresponds to $\sigma_{gas}\propto f_{cyl}^{-1}$ meaning that velocity dispersion mimics the $\alpha_{vir}$ and anticorrelates with linemass for filaments experiencing a higher pressure as in Cases 3 and 4. The bold black and blue lines are shown on the middle and bottom panels merely for comparative purposes.}
\end{figure}
The apparent dependence of filament evolution on external pressure is evident from the forgoing and so it is instructive to examine the impact of external pressure on the filament linemass. The top panel of Fig. 8 shows the time-variation of the mean filament linemass; just to remind our readers, $f_{cyl} = \frac{M_{l}}{M_{l_{crit}}}$. The linemass, ${M_{l}}$, is calculated by estimating the average density and filament-radius at different points along its length, i.e,
\begin{equation}
  M_{l}(l) \equiv \rho(l) r^{2}(l).
\end{equation}
To calculate the radius of the filament at different points along its length, gas particles were first segregated by eliminating the externally confining {\small ICM}particles. The distance of a particle from the spine of the filament was taken to be its radius. Thus $\rho(l)$, and $r(l)$ are respectively, the mean density and the mean radius at different points along the length of the filament. Finally, the mean linemass, $M_{l}=\frac{1}{N_{int}}\sum_{l=1}^{N_{int}} M_{l}(l)$, where $N_{int}$ is the number of points along filament length. The number of bins along the length of the filament, $N_{int}$, was calculated using the Sturges criterion for optimal bin-size (Sturges 1926).\\\\
The time variation of the linemass clearly mimics the behaviour of the underlying density-profile of the filament. Observe that the filament in each realisation is initially sub critical but the peaks in the magnitude of $f_{cyl}$ become more pronounced with higher external pressure. At a relatively low external pressure, e.g., $\frac{P_{ext}}{k_{B}}\le 10^{5}$ K cm$^{-3}$ (Cases 1 and 2),  however, the filament linemass shows only a gradual enhancement with a steep rise only towards the terminal epoch when isolated collapsing regions occur in the filaments. This behaviour is seen even when the filament is allowed to evolve with an isothermal equation of state (Case 5). \\\\
It is interesting to note that although the filaments at external pressure $\ge 10^{6}$ K cm$^{-3}$ acquire supercriticality fairly soon in their evolutionary sequences, neither do 
they enter into a runaway collapse mode, nor fragment. Instead, filaments in these Cases (3 and 4) buckle. In fact, since the filaments in these latter two cases accrete gas rapidly (see the accretion velocity, $V_{acc}$), they become super - critical on a relatively short timescale in their evolutionary sequence. By contrast, when the natal filament accretes gas relatively slowly (as in Cases 1 and 2), the increase in filament linemass is gradual, but slow. \emph{This observation suggests that the linemass alone is not a reliable proxy for understanding filament evolution.}
\begin{table}
\centering
\caption{Spearman rank coefficient ($r_{s}$) for respective data plotted on the central and lower panels of Fig. 8.}
\begin{tabular}{|l|c|c|}
\hline
Case No. & $r_{s}$ & $r_{s}$ \\
\hline
1 & 0.23 & 0.09\\
2 & 0.65 & 0.19\\
3 & -0.73 & -0.61\\
4 & -0.43 & -0.30 \\
\hline
\end{tabular}
\end{table}
\subsubsection{Virial parameter and filament linemass} The central panel of Fig. 8 shows a
correlation between the Virial parameter, $\alpha_{vir}$, and the filament linemass. As before we prefer to show the quantity $f_{cyl}$ in this plot. The Virial parameter was calculated here as -
\begin{equation}
 \alpha_{vir}=\frac{(E_{turb} + E_{therm})}{E_{grav}},
\end{equation}
where $E_{grav}$, $E_{therm}$, and $E_{turb}$ are respectively the components of energy due to self-gravity, thermal energy, and turbulent energy. As before, particles representing the {\small ICM} in the background were first eliminated. The respective energy terms were then calculated as follows; $E_{therm} = M_{fil}a_{avg}^{2}$, $E_{turb} = M_{fil}\sigma_{gas}^{2}$ and $E_{grav} = \frac{GM_{fil}^{2}}{\mathrm{0.5{\small FWHM}_{fil}}}$, so that the equation above becomes -
\begin{equation}
\alpha_{vir} = \frac{0.5\mathrm{{\small FWHM}_{fil}}(a_{avg}^{2} + \sigma_{gas}^{2})}{GM_{fil}},
\end{equation}
where $a_{avg}$ and $\sigma_{gas}$ are respectively, the sound speed corresponding to the density averaged mean gas temperature of the filament and the velocity dispersion of gas in the filament. Note that this definition of the Virial parameter is based on comparing the relative strength of the thermal support and turbulent support against the self-gravity of the filament. Although this definition is slightly different from the one proposed by Fiege \& Pudritz (2000) which explicitly invokes the filament-linemass, it carries the same essence nevertheless. As before, various points on this plot correspond to the filament linemass and its respective Virial parameter at different epochs of its evolution. Interestingly, this plot also shows that Virially bound ($\alpha_{vir} < 1$) filaments can indeed remain sub critical and in fact, core-formation can even commence in such filaments. \\\\
In a study of \emph{Herschel} filaments, Arzoumanian \emph{et al.} (2013) observed that $\alpha_{vir}$ was approximately inversely proportional to the filament linemass for sub critical filaments. Although the correlation observed here between these two quantities is similar to that observed by Arzoumanian \emph{et al.}, there is a dearth of data points for super critical filaments because the realisations were terminated soon after the filaments became super critical and locally collapsing regions appeared. Furthermore, as reflected by the Spearman rank coefficient for these data, as listed in column 2 of Table 2, the two quantities are weakly correlated in the first two cases where $P_{ext}/k_{B}\lesssim 10^{5}$ K cm$^{-3}$, and anticorrelate at higher pressure in the remaining two cases. Note that since $\alpha_{vir}$ is a direct proxy of the strength of turbulence, it must mimic the magnitude of velocity dispersion, $\sigma_{gas}$. \\\\
Similarly, the bottom panel of Fig. 8 shows the correlations between the velocity dispersion, $\sigma_{gas}$, and the filament linemass. As in the earlier plot, the dichotomy in the correlation due to the difference in filament evolution as a function of external pressure, is also visible here. At lower pressures, the correlation is similar to the observations of Arzoumanian \emph{et al.}, though slightly steeper. This similarity at lower external pressure could mean that radial oscillations of the filament add to the net energy budget of the filament so that $\sigma_{gas}$ increases as the linemass also increases. Indeed, there is evidence of radial contraction towards the {\small DR21} filament (Schneider \emph{et al.} 2010), which was also one of the objects observed by Arzoumanian \emph{et al.}. More recently, Keown \emph{et al} (2019) estimated the external pressure around {\small DR 21} to be on the order of ~10$^{4}$ K cm$^{-3}$ - 10$^{5}$ K cm$^{-3}$, similar to Cases 1 and 2 in this work. For $P_{ext}/k_{B}\gtrsim 10^{6}$ K cm$^{-3}$, however, the correlation indeed mimics the $\alpha_{vir} - f_{cyl}$ correlation, but is manifestly dissimilar to the observations of Arzoumanian \emph{et al.} Buckling of the filament followed by its rupture is, on the other hand, dissipative which could cause $\sigma_{gas}$ to anticorrelate with $f_{cyl}$ at higher external pressure (black line in the plot on lower panel of Fig. 8). As before, the Spearman rank coefficient for these data, as listed in column 3 of Table 2, reflects this dichotomy. We, however, note that the dichotomy observed in these two plots could be merely owing to the dearth of data corresponding to higher $f_{cyl}$ in this work. Irrespective of these differences, we find that the correlations observed for filaments in the field have their origins in the early stages of filament evolution itself. Data for Case 5 have not been shown in these latter two plots because the velocity dispersion in this case is approximately similar to that observed in Case 2 as is indeed visible from the plot on the lower panel of Fig. 3.
\subsubsection{External pressure and the filament column density}
Contraction of the filament while it continues to accrete gas, and the subsequent enhancement of the central density suggests a possible correlation between $\sigma_{gas}$ and the filament column density, $N_{H_{2}}$. The plot on the upper left-hand panel of Fig. 9 explores this possibility. This plot shows that the respective data are in general inconsistent with a single power-law. The Spearman rank coefficient for the respective cases listed in Table 3 shows that the two quantities are weakly anticorrelated in the first case and moderately correlated in the remaining cases. This difference suggests the position of a filament in the $\sigma_{gas} - N_{H_{2}}$ plane depends exclusively on its evolutionary sequence. The plot between the linemass, $M_{l}$, and $N_{H_{2}}$ on the upper right panel of Fig. 9 leads to a similar conclusion. The filament linemass moderately correlates with the column density in the first two cases and weakly anticorrelates in the remaining cases. The absence of a strong correlation between $\sigma_{gas}$, $N_{H_{2}}$, and $M_{l}$ once again underscores the complex nature of filament evolution and the apparent role of ambient pressure in influencing it. \\\\
\begin{table}
\centering
\caption{Spearman rank coefficient ($r_{s}$) for respective data plotted on the upper left and right hand panels of Fig. 9.}
\begin{tabular}{|l|c|c|}
\hline
Case No. & $r_{s}$ & $r_{s}$ \\
\hline
1 & -0.27 & 0.39\\
2 & 0.64 & 0.74\\
3 & 0.29 & -0.53\\
4 & 0.61 & -0.33 \\
5 & 0.28 & 0.34 \\
\hline
\end{tabular}
\end{table}
Similarly, the lower left-hand panel of Fig. 9 shows the column density of the filament as a function of the external pressure and indicates that the filaments are likely to have a roughly uniform column density over a wide range of external pressure. The continuous black line as before represents the median, here of the column density of the filament in each realisation which evidently is only weakly correlated to the external pressure. This result is not only counter-intuitive, but also inconsistent with the prediction of Fischera \& Martin (2012) who suggested a much steeper dependence of the column density on external pressure ($N_{H_{2}}\propto P_{ext}^{0.5}$). This weak dependence of column density on external pressure is similar to that observed between $\mathrm{{\small FWHM}_{fil}}$ and $P_{ext}$ earlier in the upper - panel of Fig. 3. It is to be noted that inference about both these correlations has been drawn on the basis of respective median values, although significant variation in the filament width and its column density can be seen over the course of its evolution. The scale of variation of course depends on the ambient environment. \\\\
\begin{figure*}
  \vspace*{1pt}
   \mbox{\includegraphics[angle=270,width=0.48\textwidth]{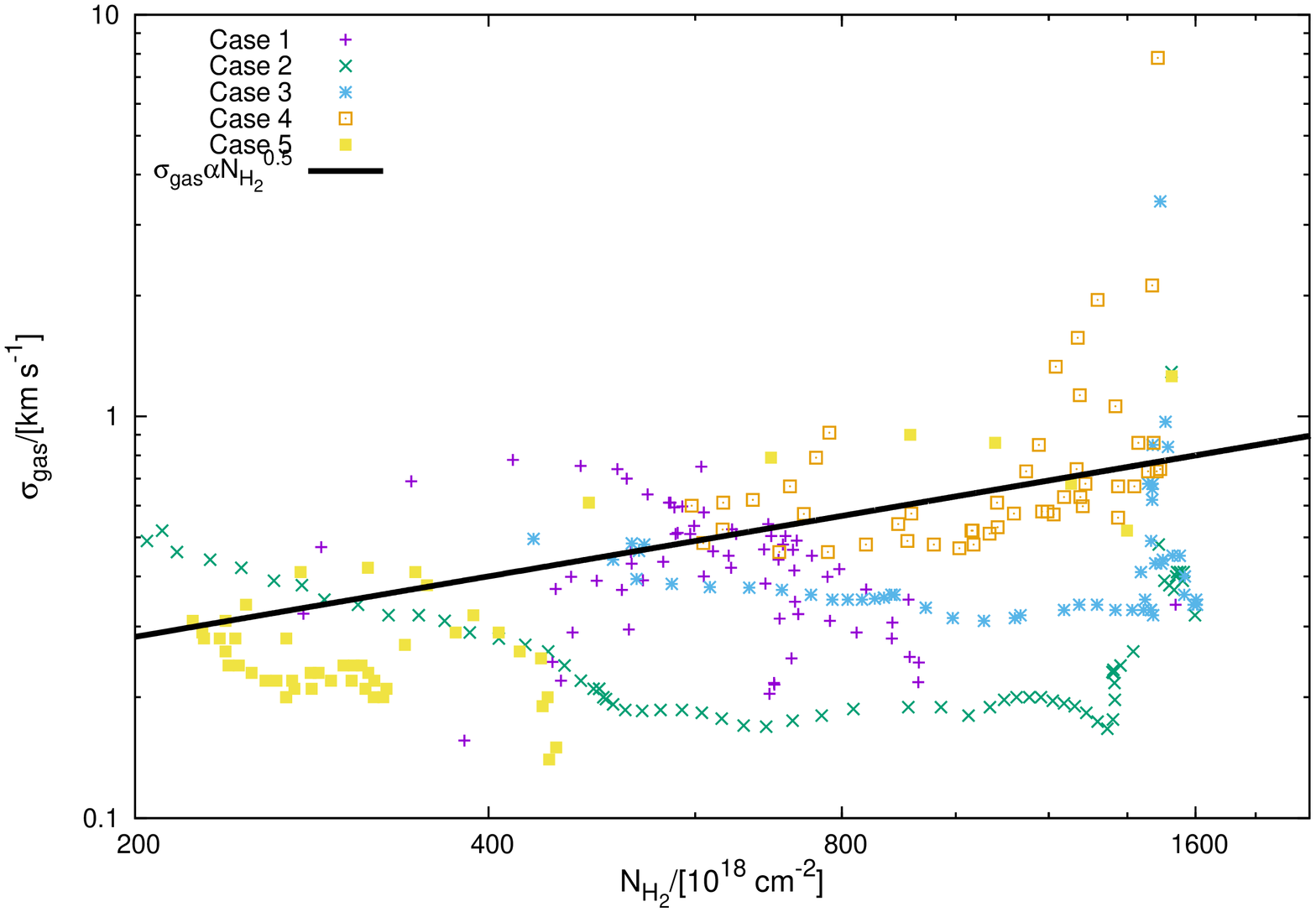}
         \includegraphics[angle=270,width=0.48\textwidth]{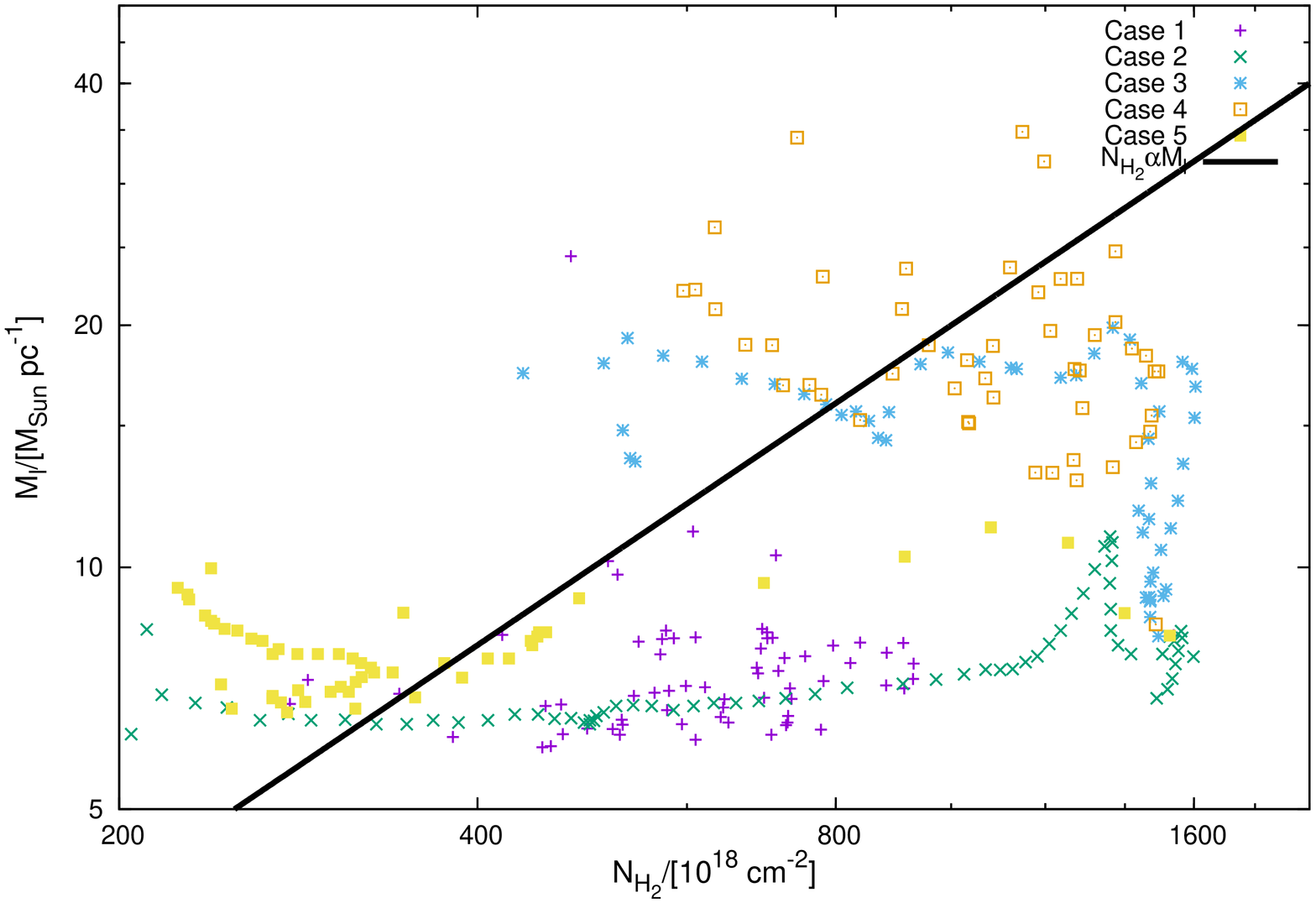}}
   \mbox{\includegraphics[angle=270,width=0.48\textwidth]{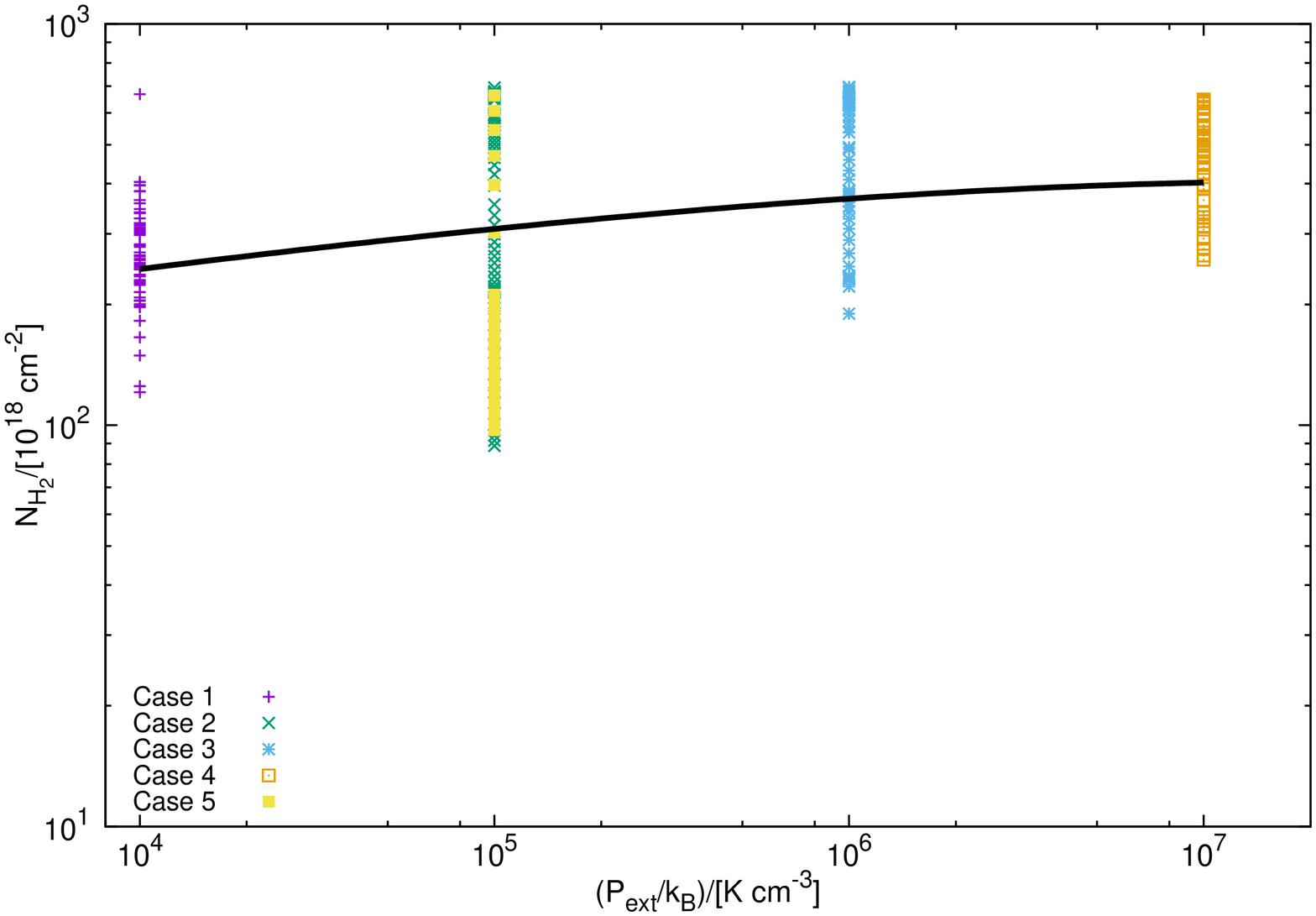}
         \includegraphics[angle=270,width=0.48\textwidth]{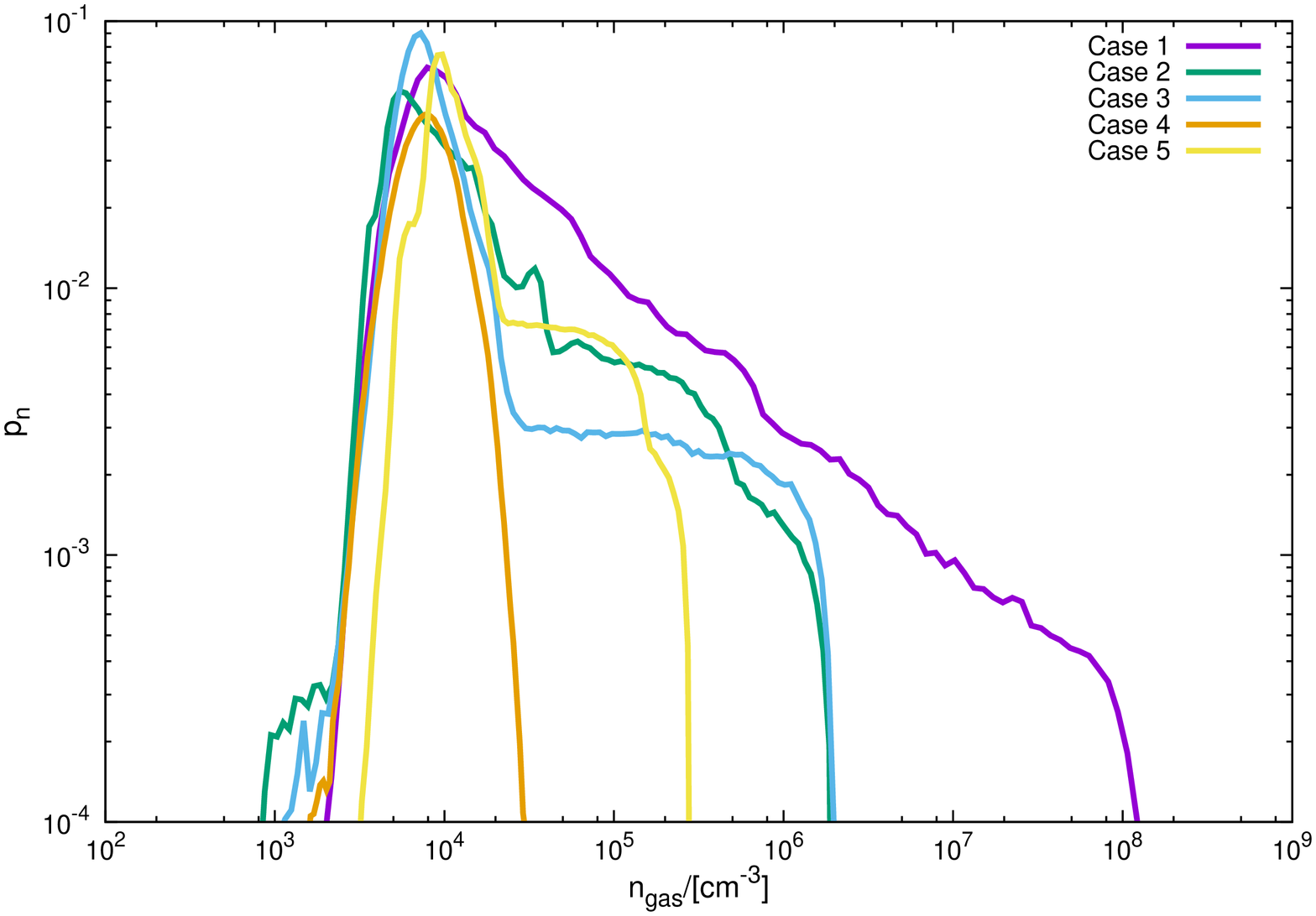}}
  \caption{Correlation between the velocity dispersion and the column density, and that between the linemass and the column density has been shown respectively on the upper left, and right hand panels. As before, the bold black line in either plot has been shown for comparative purpose only. The column density as a function of the external pressure and the density {\small NPDF} for the filament in each realisation has been shown on respectively the lower left, and the right-hand panel.}
\end{figure*}
Finally, the lower right-hand panel of Fig. 9 shows the probability distribution function ({\small PDF}) of volumetric density for the filament in each realisation at the epoch just before termination. Note that the density {\small PDF} at external pressure $\ge 10^{6}$ K cm$^{-3}$ is significantly different from the one at lower pressure. In the latter case, the {\small PDF} has a well-defined power-law tail extending to higher densities but, by contrast, this tail vanishes for $P_{ext}/k_{B}\ge 10^{6}$ K cm$^{-3}$. The {\small PDFs} in former cases are qualitatively similar to the extinction (column density) {\small PDFs} deduced by Kainulainen \emph{et al.} (2009) for several star-forming filaments in the local neighbourhood. In their work, quiescent filaments exhibited roughly lognormal extinction {\small PDFs} similar to those seen here for $P_{ext}/k_{B}\ge 10^{6}$ K cm$^{-3}$. This qualitative similarity, however, does not automatically mean that quiescent filaments in the Kainulainen \emph{et al.} survey also experience similarly large external pressures since other environmental factors in the local neighbourhood such as large-scale shear could also inhibit star-formation (e.g. Anathpindika \emph{et al.} 2018). The density {\small PDFs} observed here are consistent with those suggested by (Vazquez-Semadeni 1994; Padoan \emph{et al.} 1999) for a turbulent {\small MC} in general. Similarly, Anathpindika \emph{et al.} (2017) showed that clouds confined by a large external pressure  develop a density {\small PDF} devoid of a power-law tail.\\\\
We have seen previously in the upper - panel of Fig. 3 and on the lower left - hand panel of Fig. 9 that the filament width, {\small FWHM}$_{\mathrm{fil}}$, and its column density, both vary weakly with external pressure. Naturally then the filament width and the column density must themselves be weakly correlated as is indeed visible from the plot in Fig. 10. In fact, this plot is consistent with a similar plot presented by Arzoumanian \emph{et al.} (2011), albeit there is a dearth of super-critical filaments in our data. \emph{Nevertheless this plot supports the suggestion of an approximately uniform filament width on the order of $\sim$0.1 pc, at least for conditions similar to those in the Solar neighbourhood (as in Cases 1 and 2). While filaments at higher pressure (as in Cases 3 and 4) exhibit a similarly weak correlation, from this plot we also infer that they have a slightly greater width. }
\begin{figure}
\vspace{1pt}
\includegraphics[angle=270,width=0.48\textwidth]{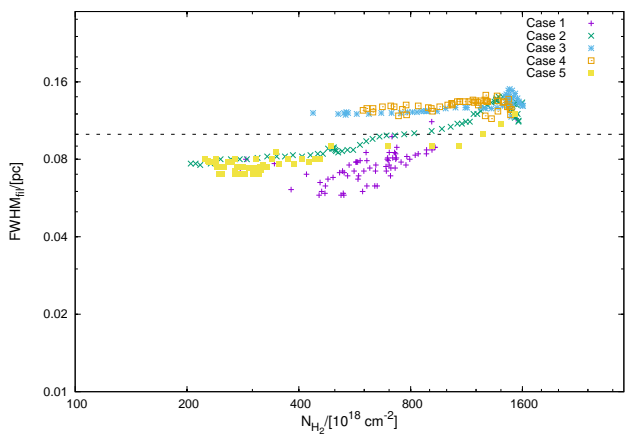}
\caption{The {\small FWHM}$_{\mathrm{fil}}$ of filaments as a function of their column densities. The dashed horizontal line represents {\small FWHM}$_{\mathrm{fil}}$ = 0.1 pc.}
\end{figure}

\subsubsection{Magnitude of external pressure and the appearance of velocity substructure} 
The various panels of Fig. 11 show the rendered images of the cross-section of the filament. Colours on this plot represent the local strength of the velocity field as indicated by the colour bar at the bottom of this figure. In the interest of brevity, we choose to call these plots the velocity rendered position-position images. The image on each panel corresponds to the terminal epoch of the filament. That the natal filament in each case is composed of more than one velocity component along its axis is evident from these images. The observed multiplicity of velocity components along the filament-axis in these plots supports the thesis that the filament is indeed composed of a bundle of multiple velocity coherent structures, often referred to as \emph{fibres} in literature (e.g., Hacar \& Tafalla 2011, Hacar \emph{et al.} 2013). \emph{The velocity rendered {position - position} images shown here lend credence to the proposition that fibres are a natural byproduct of the filament-formation process during which material is steadily accreted from the turbulent medium}.    
\begin{figure}
\vspace{1pt}
\centering
   \includegraphics[angle=0,width=0.48\textwidth]{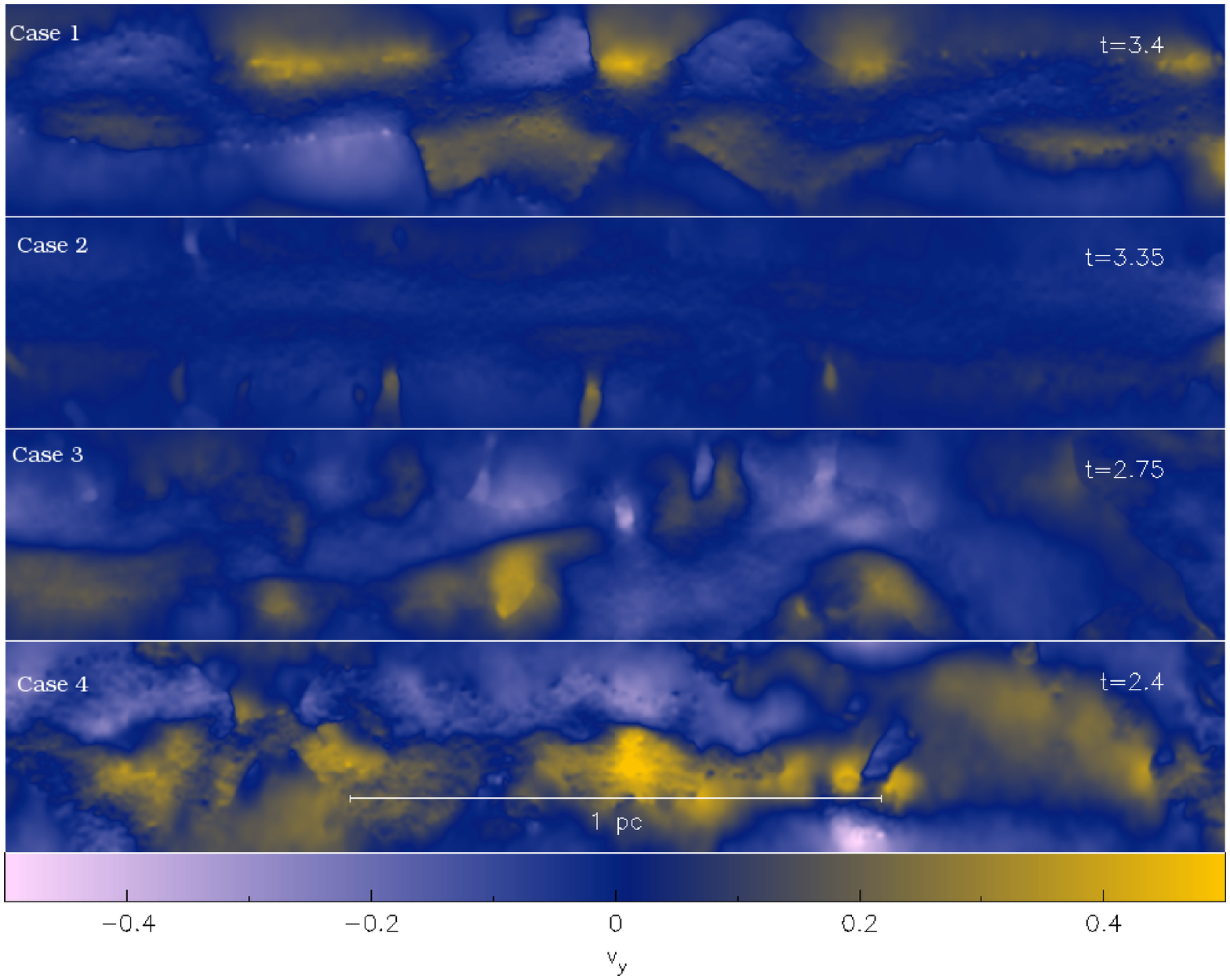}
\caption{\textbf{Velocity rendered position - position images} for respective realisations. Multiplicity of the $V_{y}$ component (in units of km s$^{-1}$) on these plots indicates that the filament is not a singular entity, but composed of several velocity coherent structures. Each plot was made at the terminal epoch of a realisation with time in units of Myr marked in the top right-hand corner of each panel. }
\end{figure}
\begin{figure*}
\begin{subfigure}{160mm}
  \vspace*{1pt}
  \centering
      \mbox{\includegraphics[angle=270,width=0.48\textwidth]{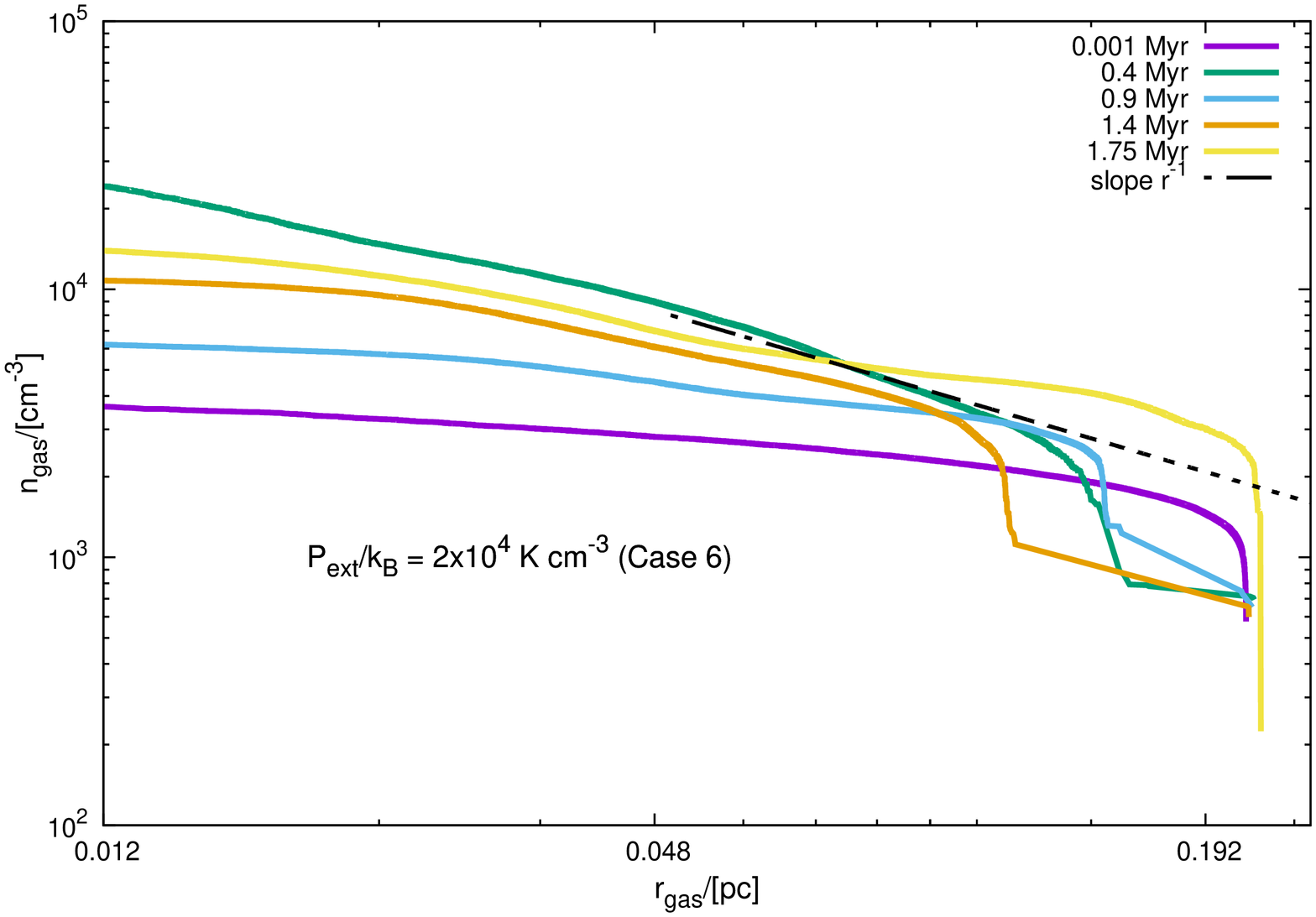}
            \includegraphics[angle=270,width=0.48\textwidth]{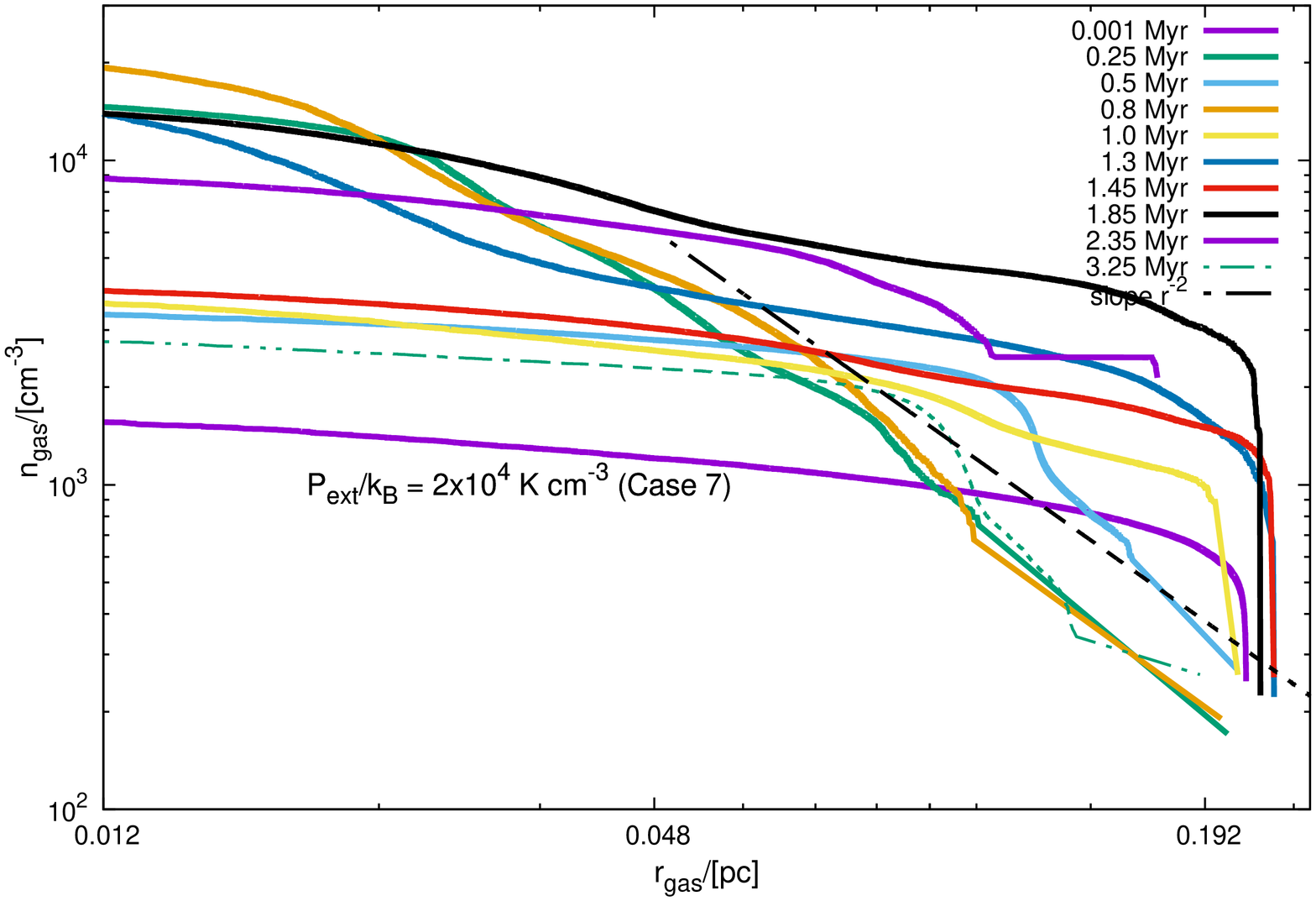}}
  \caption{Density profiles similar to those shown in Fig. 2, but now for $f_{cyl}$ = 0.7 (Case 6) and 0.3 (Case 7).}
\end{subfigure}
\begin{subfigure}{160mm}
\vspace*{1pt}
\centering
   \includegraphics[angle=270,width=\textwidth]{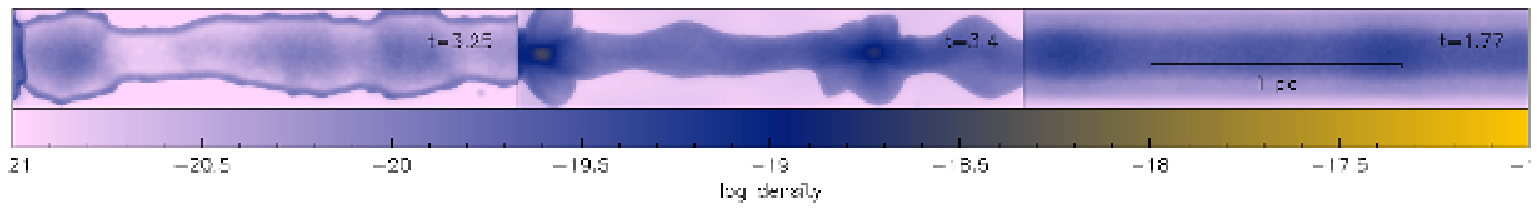}
\caption{Rendered density images showing the terminal epoch of the filament for $f_{cyl}$ = 0.3 (Case 7), 0.5 (Case 1) and 0.7 (Case 6), respectively. As before, time in units of $Myrs$ is marked in the top right-hand corner of each image. Notice that a filament with a higher initial linemass tends to fragment earlier.}
\end{subfigure}
   \caption{As with Case 1 the filament in Cases 6 and 7 experiences $P_{ext}/k_{B}$ = 2$\times 10^{4}$ K cm$^{-3}$. }
\end{figure*}
\subsection{Effect of varying the initial linemass ($f_{cyl}$ = 0.3 and 0.7)}
That the variation in the linemass of the initially sub-critical filament does not qualitatively alter the evolution of a filament for a given external pressure is evident from the plots shown in Figs. 12 (a) and (b). The respective density profiles on the left and right-hand panels of Fig. 12 (a) are similar to those shown in Fig. 2 (resp. panels c, d and e), where the density profile is relatively shallow and varies as $r^{-1}$. Radial contraction of the filament, the subsequent oscillations of the filament and the relatively shallow density profile, are common features irrespective of the magnitude of $f_{cyl}$. Increasing the linemass in this work does not significantly alter the peak density acquired by the filament as it remains on the order of a few times $10^{4}$ cm$^{-3}$, similar to the respective plots in Fig. 2.\\\\
Although we do not start with an initially super-critical filament in any case, work by Inutsuka \& Miyama (1997) shows that an initially super-critical filament contracts rapidly without ever fragmenting. This trend is visible from the plots shown on the rendered images of Fig. 12 (b), i.e., the contraction timescale for the filament is the shortest for the case with $f_{cyl}$ = 0.7 (right hand panel), and even the fragments are not clearly distinguishable as in other realisations with lower linemass. Since a filament does not ever enter a runaway collapse mode, however, it is unlikely the peak central density would change much even if our model filament were initially super-critical. Furthermore, as with the density profiles seen in rest of the cases, the slope of the density profile in outer regions of the filament in this case also lies between $r^{-2}$ and $r^{-1}$. Note that the dashed black line only reflects the slope of the density profile at the epoch when the filament has acquired its peak density. Nevertheless, fragments  are pinched in the filament with $f_{cyl}$ = 0.5, but they are broad in the sub-critical filament as is visible from the images on the left and central panels in Fig. 12 (b). In fact, the tendency of merger between fragments in the initially sub-critical filament is also evident from the image in the left-hand panel of Fig. 12 (b). At this stage it is unclear if the fragments in a sub-critical filament will ever go on to sub-fragment. \emph{Results discussed here show that variation in the linemass for a given external pressure does not qualitatively alter the evolutionary diagnostics of an initially sub-critical  filament, besides shortening the timescale of contraction for higher linemass. }
\section{Discussion}
There is mounting evidence from recent observations of {\small MCs} that suggests external pressure must affect the physical properties of {\small MCs} and as a direct consequence, their ability to form stars (e.g., Hughes \emph{et al.} 2013, Meidt \emph{et al.} 2013). In other words, the external pressure must somehow influence the processes that convert diffuse gas in a {\small MC} into putative star-forming regions that usually appear filament-like. Analytic calculations predict that filaments should be more centrally condensed ($N_{H_{2}}\propto P_{ext}^{0.5}$), and diminishing width ($FWHM\propto P_{ext}^{-0.5}$), with increasing external pressure (Fischera \& Martin 2012). Numerical simulations of individual clouds, on the other hand, show clearly that the evolution of a cloud confined by a relatively large external pressure is much more complex than analytic calculations predict (Anathpindika \emph{et al.} 2017). Also, unlike the relatively simple analytic predictions,  
the radial density profiles shown in Fig. 2 show radial oscillations of the filament. \\\\
As was shown in the case of a purely thermally supported filament by Anathpindika \& Freundlich (2015), a turbulence-supported filament also contracts in a quasistatic manner to acquire a peak central density, but never enters runaway collapse.
Plots on various panels of Fig. 2 and in Fig. 12(a) collectively highlight that irrespective of the initial filament linemass, and for typical magnitudes of external pressure, the density profile at large radii is relatively shallow (between $r^{-2}$ and $r^{-1}$), and therefore inconsistent with the profile of an isothermal cylinder, i.e., $n_{gas}\propto r^{-4}$ (Ostriker 1964). The density profiles seen here are in general consistent with those reported observationally for typical filamentary clouds (e.g., Malinen \emph{et al.} 2012, Palmeirim \emph{et al.} 2013), who reported profiles of the type $r^{-2}$. In fact, even the density profile for the purely isothermal realisation varies as $r^{-1}$, unlike the prediction of the Ostriker model, which may be attributable to the turbulence injected by the accreting gas and the fact that the model filament is itself initially partially supported by turbulence. \emph{Density profiles deduced in this work show that filaments need not be threaded with a magnetic field to produce a shallow density profile as is often suggested (e.g., Fiege \& Pudritz 2000, Federrath 2015, Hennebelle \& Inutsuka 2019)}. \\\\
The injection of turbulence within a filament, and the additional support it provides against self-gravity is the key here, as is reflected by the weak dependence of the $\mathrm{{\small FWHM}_{fil}}$ on the magnitude of $P_{ext}$. Indeed, the trend visible in the upper-panel of Fig. 3 suggests that the filament width increases slightly with increasing pressure, though it is roughly on the order of $\sim 0.1$ pc for external pressure similar to that in the Solar neighbourhood (Cases 1 and 2). This result is largely consistent with the conjecture of a \emph{universal filament width} based on estimates for \emph{Herschel} filaments across a number of nearby clouds (e.g., Arzoumanian \emph{et al.} 2011, 2013, 2019; Palmeirim \emph{et al.} 2013; Koch \& Rosolowsky 2015). These estimates of filament width are also corroborated by other independent observations of filaments (e.g., Salji \emph{et al.} 2015). In view of the data shown in the plot on the upper-panel of Fig. 3, however, we would prefer to restrict the universality of filament width to filaments in the Solar neighbourhood only where the ambience is unlikely to vary much. Analytic  predictions,  on the other hand, favour thinner filaments with increasing magnitude of external pressure (Fischera \& Martin 2012). This expected behaviour is based on the assumption of hydrostatic equilibrium in the analytic calculations and therefore does not account for the conditions of the ambient environment. \\\\
Numerical simulations modelling structure-formation in turbulent gas often characterise the filament width as the length scale over which turbulence makes a transition to the subsonic domain (e.g., Smith \emph{et al.} 2014, Federrath 2015, Ntormousi \emph{et al.} 2016). According to analytic estimates, this length is the scale over which filaments acquire a quasi equilibrium structure at an external pressure on the order of a few times 10$^{4}$ K cm$^{-3}$ (Fischera \& Martin 2012). An alternative explanation to the filament width relies on the thermodynamic equilibrium achieved by the filament with the external medium and thus ties the characteristic width to the local thermal Jeans length (e.g., Anathpindika \& Freundlich 2015; Hocuk \emph{et al.} 2016). These arguments, however, ought to be revisited in view of the plots on the various panels of Fig. 3. Collectively, these plots suggest that the filament width must be determined by the interplay between self-gravity of the filament, the external pressure, the local sound speed and the velocity dispersion. \\\\
For a filament in approximate pressure balance we have -
\begin{equation}
P_{ext} + P_{grav}\equiv a_{eff}^{2}\rho_{gas},
\end{equation}
where $P_{grav}$, is the pressure due to self-gravity of the filament, $a_{eff}^{2}\equiv a_{0}^{2} + \sigma_{gas}$, is the effective sound speed, $\rho_{gas}$, is the average gas density, and $P_{ext}$, as before, is the external pressure. For a filament of mass, $M_{fil}$, and radius, $R_{fil}$, the pressure due to self-gravity is then merely $P_{grav}\equiv\frac{2GM_{fil}^{2}}{R_{fil}^{4}}$, so that
\begin{equation}
R_{fil}=\frac{G^{0.25} M_{fil}^{0.5}}{(a_{eff}^{2}\rho_{gas} - P_{ext})^{0.25}}.
\end{equation}
Figure 13 shows the filament width \textbf{($\sqrt{2\ln 2}R_{fil}$)} as a function of average gas density for different choices of external pressure and effective sound speed used in the realisations discussed above. These fiducial calculations were performed for two choices of filament mass, $M_{fil}$. Evidently, for the physical conditions and range of densities typically found in star-forming regions in the local Solar neighbourhood, the filament width is generally on the order of $\sim$0.1 pc. In fact, the filament width converges to this value for $P_{ext}/k_{B} > 10^{4}$ K cm$^{-3}$ at gas densities on the order of few times $10^{4}$ cm$^{-3}$. Filaments of smaller width are likely in denser environs where the average density itself is significantly higher. Indeed, smaller filament widths (of up to $\sim 0.03$ pc) for sub-structures, which may loosely be referred to as \emph{fibres}, along the main filament of the infrared dark cloud ({\small IRDC}) {\small G035.39 - 00.33} have also been reported (e.g. Henshaw \emph{et al.} 2017). Though this small width could be an artefact of data filtering, {\small IRDCs} are typically 2-3 orders of magnitude denser than the nearby clouds (e.g., Perault \emph{et al.} 1996), so filaments with higher average density could indeed be thinner. We briefly observe thinner filaments (see images in Fig. 5(a)) in realisations with $P_{ext}/k_{B}\ge 10^{6}$ K cm$^{-3}$, but these are unstable and buckle.  \emph{The plot in Fig. 13 shows that filament widths are likely to be on the order of $\sim$0.1 pc for physical conditions typically found in the local Solar neighbourhood, but filaments in relatively dense gas could be narrower.} 
\begin{figure}
  \vspace*{1pt}
  \includegraphics[angle=270,width=0.5\textwidth]{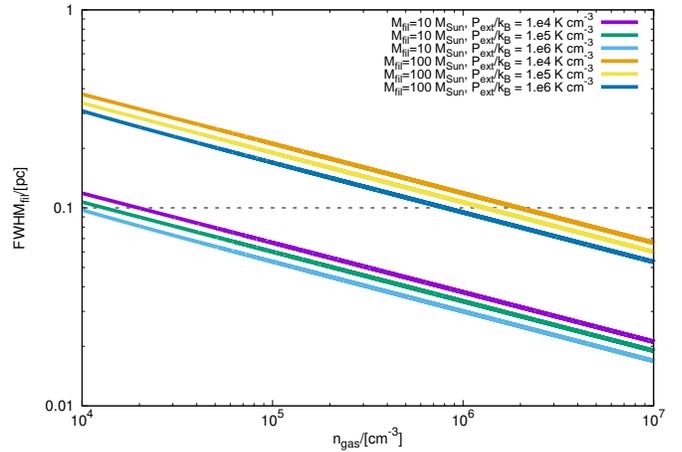}
  \caption{The filament width as a function of the average gas density for different choices of external pressure as predicted by Eq. (10).}
\end{figure}
\subsection{Filament fragmentation and the size of fragments}
While a majority of the prestellar cores detected in nearby filaments have sizes smaller than the width of their natal filament, cores in some filaments such as the Musca filament are broad. Although the filament in the realisations discussed here fragments due to the so-called \emph{sausage-like} deformation instability, pinched fragments are observed only in Case 1 (see images in Fig. 5 (a)), where the external pressure is, $P_{ext}/k_{B}\sim 10^{4}$ K cm$^{-3}$. Fragments in the rest of the realisations are broad and roughly spherical. In their work on filament fragmentation, Inutsuka \& Miyama (1997) observed the prevalence of sausage-type instability in initially transcritical filaments ($f_{cyl}$ = 1), confined by a relatively large external pressure. On the other hand, a radial collapse leading to a spindle was observed in the case of initially super critical filaments, without any external perturbations and irrespective of the magnitude of external pressure.\\\\
A filament assembled in a typical {\small MC} must essentially start from a sub critical phase and then gradually accrete gas. Unlike Inutsuka \& Miyama (1997), we therefore began with a sub critical or at best, a transcritical filament. Nevertheless, there is qualitative similarity to a large extent between our results and inferences drawn by the former authors. Like them, we also observe the dominance of sausage-type instability leading to spherical fragments in a filament at relatively high external pressure. Broad, spherical fragments are also observed when the filament is initially sub critical even if the external pressure is relatively small (Case 7). By contrast, pinched fragments are observed at low external pressure ($\sim 10^{4}$ K cm$^{-3}$), but only when the filament is initially transcritical (Cases 1 and 6). Furthermore, the initially transcritical filament in Case 4 soon became super critical (plot on the upper-panel of Fig. 8), but unlike the observation by Inutsuka \& Miyama (1997), the filament did not rapidly contract radially. Instead, it buckled to produce composite fibres. \emph{Evidently, the magnitude of external pressure plays a crucial role in the eventual fate of the filament. Filaments confined by a relatively small pressure such as that resulting from the convergence of slow moving gas flows are more likely to form pinched cores. By contrast, relatively large external pressure tends to eventually disrupt the filament which could possibly help reconcile the poverty of star-formation in high pressure environs (see e.g., Longmore \emph{et al.} 2013).} \\\\
Observations of several nearby filaments have shown the existence of cores smaller than the average filament width (e.g., Palmeirim \emph{et al.} 2013, Marsh \emph{et al.} 2014), while those in the {\small B213} complex of the Taurus {\small MC} (e.g., Tafalla \& Hacar 2015), and the Musca filament (e.g., Hacar \emph{et al.} 2016), are broad. In a more recent numerical work Gritschneder \emph{et al.} (2017) showed that a filament with initially small sinusoidal perturbations fragmented to form cores as these perturbations amplified. This behaviour is indeed that expected in the case of the deformation instability identified by Nagasawa \emph{et al.} (1987). By contrast, arguing about the dependence of filament evolution on its linemass Heigl \emph{et al.} (2019) showed that the isothermal contraction of an initially transcritical filament confined by pressure comparable to that in the Solar neighbourhood, led to pinched cores whereas an initially sub-critical filament formed broad cores.  Collectively, these recent studies emphasise the importance of filament linemass. Furthermore, unlike the analytic result by Nagasawa (1987), it is unclear from these latter numerical works if the initial filament radius was smaller than its scale height. \emph{Instead, we argue that the magnitude of external pressure is crucial towards determining the eventual fate of a filament. We have seen (in Cases 1 and 6) that pinched fragments form in a transcritical filament confined by a relatively small external pressure (on the order of 10$^{4}$ K cm$^{-3}$ - 10$^{5}$ K cm$^{-3}$). For such small values of external pressure, however, broad, spherical fragments form if the filament is initially sub-critical. Also, irrespective of the initial linemass of the filament, broad spherical fragments result if the external pressure is relatively large.}
\subsection{Other diagnostic properties of filaments}
Several surveys of filaments in nearby {\small MCs} have also shown a direct correlation between the velocity dispersion, $\sigma_{gas}$, within a filament, its linemass and the column density (e.g., Arzoumanian \emph{et al.} 2013). These observed correlations lend support to the suggestion that accretion injects turbulence into the filament while also increasing its column density. The injected turbulence supports the filament against self-gravity and is therefore an important factor in determining its characteristic width. The plot on the lower panel of Fig. 8 shows the dichotomy driven by external pressure in the correlation between filament linemass and velocity dispersion. The turbulent velocity is smaller in filaments with a higher linemass, and a relatively large external pressure ($\gtrsim 10^{6}$ K cm$^{-3}$, viz., Cases 3 and 4). For filaments experiencing a smaller external pressure ($\sim 10^{4}$ K cm$^{-3}$ in Case 1), the velocity dispersion is fed by the radial oscillations of the filament and therefore tends to increase with increasing linemass. Data points for Case 2 ($\sim 10^{5}$ K cm$^{-3}$) are a mixture as some of these are consistent with the dashed blue line while others suggest a roughly constant, or even weakly decreasing $\sigma_{gas}$ with increasing linemass. \\\\
There is no significant correlation between the filament linemass and its column density as seen in the upper right-hand panel of Fig. 9. This behaviour occurs because the filament in each realisation becomes super-critical only in the latter stages of its evolution, as can be seen in the upper-panel of Fig. 8, by which time the filament acquires approximate hydrodynamic equilibrium. As was reported by Arzoumanian \emph{et al.} (2011), the plot in Fig. 11 shows that the filament width and the column density are weakly correlated and that filaments in solar neighbourhood-like ambience are likely to have roughly uniform width on the order of $\sim$ 0.1 pc. It is no doubt true that filaments in these realisations could not be followed to still higher column densities because calculations terminated soon after sites of local collapse appeared (Note also that the middle panel of Fig.12b corresponding to Case 6, $f_{cyl}$ = 0.6, shows only the central section of the filament and not its edges where localised collapse was observed.) Although the plots in Figs. 8 and 9 lack sufficient data points corresponding to super-critical filaments so that the complete extent of these respective correlations is not fully spanned, these plots are, however, illustrative subsets of the $\sigma_{gas}$-$N_{H_{2}}$ and $M_{l}$ - $N_{H_{2}}$ correlations observed for filaments in the field. 
\subsection{Velocity coherent structure}
In the so-called \emph{fray and fragment} scenario (e.g., Hacar \emph{et al.} 2013), colliding turbulent flows in a {\small MC} generate filamentary structure. The bundles of fibres associated with individual filaments are generated due to the turbulence within the resulting filaments and are thus, a natural byproduct of this process. The rendered velocity images in Fig. 11 show that the natal filament is itself composed of multiple velocity components along its axis. Owing to their contiguity, we refer to these substructures as \emph{fibres}. They exhibit a moderately supersonic velocity gradient (with reference to the mean gas temperature of the filament when calculations were terminated) on the order of 0.4 km s$^{-1}$ pc$^{-1}$- 0.6 km s$^{-1}$ pc$^{-1}$, and could fragment further to form smaller cores. These images are consistent with the observations of the {\small L1495/B213} region in Taurus by Hacar \emph{et al.} (2013). Those data motivated the scenario of \emph{hierarchical fragmentation} of a filament wherein the major filament first fragments into smaller velocity-coherent fibres on sub-parsec scale. This behaviour is indeed seen in filaments in this work, irrespective of the magnitude of external pressure. The relatively simple realisations developed in this work show that the natal filament splits quite naturally into several constituent \emph{fibres} over the course of its evolution. \\\\
Results deduced here are consistent with those reported in some earlier works, e.g. see, Smith \emph{et al.} (2014), Moeckel \& Bukert (2015), Zamora - Avil{\' e}s \emph{et al.} (2017). Although the latter authors also caution that the apparent existence of fibres could be merely an artefact of the line-of-sight effects, our results support the idea that fibres are integrally associated with the filament formation process. In some more recent work, Clarke \emph{et al.} (2017) showed that a filament exhibited a greater propensity to form fibres when stronger turbulence was injected in to it. They also showed that with increasing strength of the injected turbulence, the filament made a gradual transition from gravity-dominated to turbulence-dominated fragmentation. By contrast, we argue here that the external pressure bears upon the morphology of filament fragmentation. 
\subsection{Limitations}
Realisations discussed in this work started with an idealised set-up which precluded us from studying the assembly and evolution of filaments and their constituent fibres in a self-consistent manner. Moreover, each set-up had only a few tens of Solar masses of gas due to which filaments could not attain the highly super-critical state, as was reflected by the dearth of data points with relatively large $f_{cyl}$ in respective correlations in Figs. 8 and 9. These respective correlations are therefore representative at best. A more realistic set-up to study the impact of ambient environment on filament formation/evolution should necessarily involve a section of the Galactic arm with some form of feedback that would inject turbulence and generate filaments in a self-consistent manner. We intend to develop this subject in future contributions.
\section{Conclusions}
With a set of relatively simple hydrodynamic realisations of a typical star-forming filament, we have showed that filament evolution is primarily governed by the interplay between self-gravity, injected turbulence, and thermal support.  Below are our principal conclusions -
\begin{enumerate}
 \item Contraction of the model filament is observed in all test cases irrespective of the magnitude of external pressure. The peak central density reached by the contracting filament, however, depends relatively weakly on the external pressure. This result suggests the gravitational state of the filament is not altered significantly by external pressure which is consistent with the earlier conclusion by Fischera \& Martin,
\item The column density of the model filament depends weakly on external pressure, which is inconsistent with the conclusion by Fischera \& Martin who suggested a much steeper correlation ($N_{H_{2}}\propto P_{ext}^{0.5}$). In fact, we do not observe significant variation of the column density over the course of evolution of our model filaments. This behaviour could be a result of the fact that the filament in each realisation was initially sub critical or at best, transcritical. Also, as noted in \S 4.4 above, owing to the nature of the set-up only a limited pool of gas was available for accretion,
 \item Results from the simulations presented here support suggestions that typical star-forming filaments are characterised by a median width on the order of $\sim$0.1 pc, especially in cases where the external pressure is comparable to that in the Solar neighbourhood. At higher pressure, $P_{ext}/k_{B}\gtrsim 10^{6}$, the parent filament appears to split into thinner component filaments (referred to as \emph{fibres} in the main text) as can be seen in the panels corresponding to Cases 3 and 4 in Fig. 5 (a). The filament width signifies the spatial scale over which self-gravity of the filament, the turbulence injected by the accreting gas, and the thermal support within a filament acquire approximate equilibrium,  
 \item Irrespective of the external pressure, the choice of equation of state or the initial linemass, filaments in this work gradually developed a density profile that has a relatively shallow slope, between $n_{gas}\propto r^{-2}$ and $n_{gas}\propto r^{-1}$ in their outer regions. This density profile is significantly shallower than the Ostriker profile ($n_{gas}\propto r^{-4}$) for an isothermal filament. Evidently, filaments need not be magnetised to  
have a shallow density profile,  
 \item External pressure is crucial towards determining the morphology of filament evolution. Irrespective of the external pressure the filaments were observed to be susceptible to the \emph{sausage}-type instability. Fragments are pinched i.e., smaller than the width of the natal filament, even in an initially transcritical filament only when the external pressure is on the order of $\sim 10^{4}$ K cm$^{-3}$, which is comparable to that found in several clouds in the Solar neighbourhood. In all other cases the fragments were observed to be approximately spherical and broad which is consistent with the findings of Inutsuka \& Miyama (1997) in general. The filament at higher external pressure ($\ge 10^{6}$ K cm$^{-3}$) buckles and becomes unstable, and  
\item Velocity coherent structures are observed in all cases, irrespective of the external pressure suggesting that fibres are associated naturally with the filament formation process. Results from simulations discussed here support the \emph{fray-and-fragment} scenario.
\end{enumerate}
The idea that ambient pressure somehow affects the evolution of molecular clouds and their ability to form stars has been gaining traction in recent years. Given the ubiquitous nature of filaments in the interstellar medium and with filaments being integral to the star formation process, it was important to investigate quantitatively the impact of external pressure on their evolution, and fragmentation. Our finding here that filaments (as in Cases 3 and 4), experiencing a relatively high pressure buckle without ever cycling much gas into the dense putative star-forming phase, will help us to reconcile the general inefficiency of star formation in high pressure environs. We also show that filaments at an intermediate pressure (in the range $10^{4}$ K cm$^{-3}$ - $10^{5}$ K cm$^{-3}$) are the most propitious for star formation. 

\section*{Acknowledgements}
This project was initiated with funds made available under the From {\small GMCs} to stars project ({\small GMCS}/000304/2014), funded by the Department of Science \& Technology, India. Simulations discussed in this work were developed using supercomputing facilities made available by WestGrid (www.westgrid.ca) and Compute Canada Calcul Canada (www.computecanada.ca). We acknowledge comments of an anonymous referee that helped clarify better some of the results discussed in this work. 

\section*{Data Availability Statement}
No valid data repositories exist as the data generated by numerical simulations discussed in this work are too big to be shared. Instead we discuss in detail the numerical methods and the initial conditions used to generate these data sets. The initial conditions file and the script of the numerical code can be made available to bona fide researchers on reasonable request.

\appendix
\section[]{Radial density profile of the filament and estimation of $\mathrm{{\small FWHM}_{fil}}$}
To estimate the filament width we first segregated the gas particles from the externally confining {\small ICM} particles. The filament was then divided in to $N_{int}$ number of sections along its length, as also described in \S 3.3 above. We then calculated the column density for each gas particle to obtain the column density profile in each section of the filament. Finally, the mean column density profile was obtained by averaging over these profiles along the length of the filament. The plot on the upper panel of Fig. A1 shows one such column density profile for the filament in Case 7 at $t\sim $2.25 Myrs. This realisation is similar to that in Case 1 and the same external pressure ($P_{ext}/k_{B} = 2\times 10^{5}$ K cm$^{-3}$). As seen in \S 3.4 above the filament evolution in this case was qualitatively similar to that in Case 1. We adopt the idealised model of a cylindrical filament to fit this density profile -
\begin{equation}
\rho(r) = \frac{\rho(r = r_{in})}{\Big[1 + \Big(\frac{r}{r_{in}}\Big)^{2}\Big]^{p/2}},
\end{equation}   
integration of which yields the corresponding profile for column density,
\begin{equation}
\Sigma(r) = X \frac{\rho_{in}r_{in}}{\Big[1 + \Big(\frac{r}{r_{in}}\Big)^{2}\Big]^{(p-1)/2}}
\end{equation}
 (Ostriker 1964), where $X$ is a numerical constant and $\rho_{in}\equiv \rho(r = r_{in})$, is the density that remains approximately uniform within the inner radius, $r_{in}$, of the filament as can be seen in the plot on the lower panel of Fig. A1. This latter plot is the same as the one on the upper panel, but now made with logarithmic axes. Observe that the column density profile of the filament in this case is relatively shallow and consistent with the profile defined by Eqn. (A2) for $p = 2$.
\begin{figure}
\vspace*{1pt}
\centering
    \includegraphics[angle=270,width=0.45\textwidth]{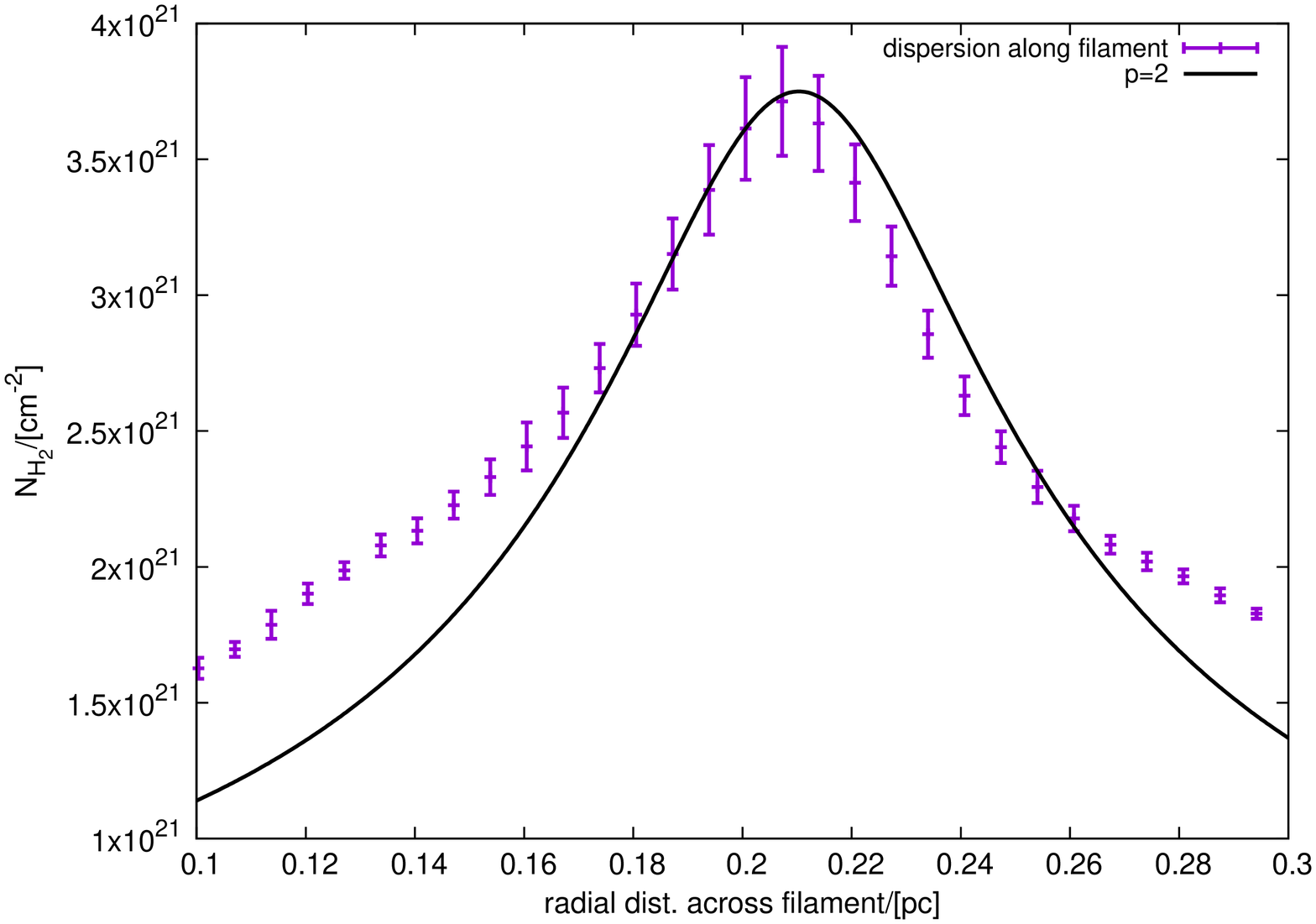}
    \includegraphics[angle=270,width=0.45\textwidth]{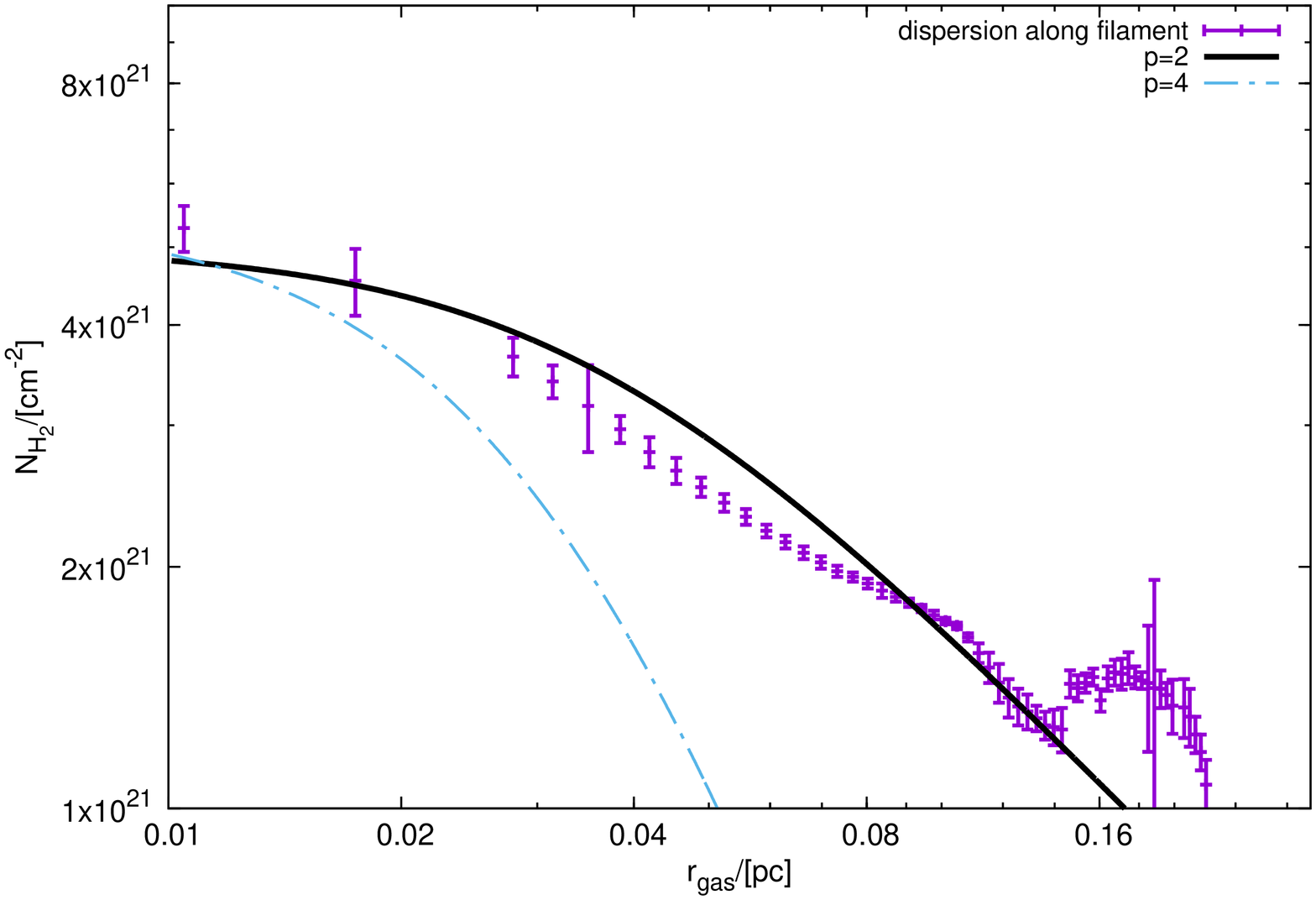}
\caption{\emph{Upper - panel :} Mean radial column density profile across the length of the filament. \textbf{The black curve overlaid on this plot corresponds to Eqn. (A2) with $p$ = 2.}   \emph{Lower - panel :} Same as that on the upper-panel but now shown as a radial profile.}
\end{figure}

\bsp

\label{lastpage}

\end{document}